\documentclass[english,aip,pop,reprint]{revtex4-2}
\usepackage[T1]{fontenc}
\usepackage[latin9]{inputenc}
\usepackage{color}
\usepackage{booktabs}
\usepackage{amsmath}
\usepackage{graphicx}
\usepackage{soul} 

\newcommand{\replace}[2]{#2}

\makeatletter

\providecommand{\tabularnewline}{\\}

\makeatother

\usepackage{babel}
\begin{document}
\title{ Do chaotic field lines cause fast reconnection in coronal loops? }
\author{Yi-Min Huang}
\email{yiminh@princeton.edu}

\affiliation{Department of Astrophysical Sciences and Princeton Plasma Physics
Laboratory, New Jersey 08543, USA}
\author{Amitava Bhattacharjee}
\affiliation{Department of Astrophysical Sciences and Princeton Plasma Physics Laboratory, New Jersey 08543, USA}
\begin{abstract}
  Over the past decade, Boozer has argued that three-dimensional (3D) magnetic reconnection fundamentally differs from two-dimensional (2D) reconnection due to the fact that the separation between any pair of neighboring field lines almost always increases exponentially over distance in a 3D magnetic field. According to Boozer, this feature makes 3D field-line mapping chaotic and exponentially sensitive to small non-ideal effects; consequently, 3D reconnection can occur without intense current sheets. We test Boozer's theory via ideal and resistive reduced magnetohydrodynamic simulations of the Boozer--Elder coronal loop model driven by sub-Alfv\'enic footpoint motions [A.~H.~Boozer and T.~Elder, Physics of Plasmas \textbf{28}, 062303 (2021)]. Our simulation results significantly differ from their predictions. The ideal simulation shows that Boozer and Elder under-predict the intensity of current density due to missing terms in their reduced model equations. Furthermore, resistive simulations of varying Lundquist numbers show that the maximal current density scales linearly rather than logarithmically with the Lundquist number.

\end{abstract}
\maketitle

\section{Introduction}

Magnetic reconnection is a fundamental mechanism that changes the topology of magnetic field lines and converts magnetic energy to plasma thermal and non-thermal energy. \citep{Biskamp2000,PriestF2000,ZweibelY2009,YamadaKJ2010,ZweibelY2016,JiDJLSY2022,PontinP2022} It is generally believed that this mechanism drives explosive phenomena in astrophysical, space, and laboratory plasmas, including solar flares, coronal mass ejections, geomagnetic substorms, and sawtooth crashes in fusion devices. 

Magnetic reconnection can be classified into two-dimensional (2D) and three-dimensional (3D) reconnection. Real-world magnetic reconnection takes place in 3D, but 2D reconnection is commonly employed as an approximation by assuming that the whole process depends only on two spatial coordinates. Two-dimensional reconnection occurs at an X-point (or X-line when extending along the direction of symmetry) where separatrices, which separate topologically different magnetic field lines, intersect. When magnetic stress builds up at the X-point prior to reconnection, a thin current sheet forms.  Then, during reconnection, the field line velocity (i.e., a velocity field that carries field lines from one time to another) diverges at the X-point. \citep{PriestHP2003} In other words, the field line is cut and rejoined with another field line at the X-point. 

Compared with 2D problems, magnetic reconnection in 3D remains a conceptual challenge, especially when topological structures, such as magnetic null points and closed magnetic field lines, are absent.\citep{Pontin2011} In this situation, all magnetic field lines are topologically equivalent, and a continuous velocity field that preserves magnetic field line connectivity can always be found.\citep{Greene1993} That raises a fundamental question: how and where does magnetic reconnection occur if all field lines are topologically identical? Several ideas have been proposed to address this question, including the general magnetic reconnection theory\citep{SchindlerHB1988,HesseS1988,SchindlerHB1991} that uses parallel voltage as a metric for 3D reconnection rate and the concept of quasi-separatrix-layers (QSLs) defined as regions with high squashing factors of the field line mapping.\citep{TitovH2002,Titov2007,TitovFPML2009} Even though 3D reconnection is a vast and ongoing research topic, most theories share one common aspect: Just like in 2D reconnection, intense thin current sheets play a critical role in 3D reconnection. 

Through a series of publications over the past decade, Boozer has advocated a paradigm shift regarding 3D reconnection.\citep{Boozer2012,Boozer2012a,Boozer2013,Boozer2014,Boozer2018,Boozer2019,Boozer2021,Boozer2022} The gist of Boozer's proposal can be described as follows. In a 3D magnetic field, neighboring magnetic field lines generically exponentiate away from each other. The field-line flow, which is Hamiltonian, becomes chaotic. If we follow two field lines initially separated by an infinitesimal distance $\delta r(0)$, in most cases the separation grows exponentially as $\delta r(\ell)=e^{\sigma(\ell)}\delta r(0)$, where $\ell$ is the distance along the field line, and $\sigma(\ell)$ is an overall (but in general non-monotonically) increasing function over distance. Boozer has argued that under the condition of large field-line exponentiation, an exponentially small non-ideal effect will completely scramble the field-line mapping, leading to fast reconnection without the necessity of intense thin current sheets.
\replace{}{Indeed, that fast reconnection can occur without intense thin current sheets is what sets  Boozer's theory apart from ``traditional'' theories, to use Boozer's terminology.\citep{Boozer2022}
} 

Recently, Boozer and Elder \citep{BoozerE2021} have proposed a simple coronal loop model to test this new paradigm. In their model, the coronal loop is enclosed in a perfectly conducting cylinder of a radius $a$ and a finite length $0\le z\le L$. The initial magnetic field is uniform and pointing along the $\boldsymbol{\hat{z}}$ direction. The magnetic field lines are line-tied to the top and the bottom boundaries, which represent the photosphere. On the top boundary at $z=L$, a time-dependent flow is imposed; all other boundaries are static. The imposed boundary flow mimics photospheric convection and gradually entangles (or ``braids'') the field lines in the coronal loop. The braided magnetic field lines eventually reconnect and release the stored magnetic energy into plasma kinetic energy and heat. 

On the surface, the Boozer--Elder model is similar to Parker's model\citep{Parker1972,Parker1988} of coronal heating, but their predictions for the scenarios of magnetic reconnection and energy release are starkly different. Parker's scenario predicts that intense thin current sheets will develop throughout the coronal loop as a consequence of field line braiding. These thin current sheets cause numerous small-scale reconnection events, or ``nanoflares'' (as Parker called them), that heat the solar corona to millions of degrees. Thin current sheets play a crucial role in Parker's scenario of coronal heating. In fact, Parker argued that the thin current sheets will become singular (i.e., take the form of Dirac $\text{\ensuremath{\delta}}$-functions) in the ideal-MHD limit when the resistivity vanishes.\citep{Parker1994} Parker's prediction of ideal singular current sheets, often dubbed the ``Parker problem,'' has remained controversial for several decades,  continuing to this day.\citep{VanBallegooijen1985,ZweibelL1987,LongcopeC1996,LongbottomRCS1998,NgB1998,CraigS2005,Low2006a,Low2007,JanseL2009,HuangBZ2009,HuangBZ2010,AlyA2010,Low2010,Low2010a,JanseLP2010,Low2011,PontinH2012,CraigP2014,CandelaresiPH2015,ZhouHQB2018,PontinH2020} 

Contrary to Parker's nanoflare scenario, intense thin current sheets play no significant role in the Boozer--Elder scenario. While Boozer and Elder also predict that electric current distribution will form thin ribbons, the current density does not become very intense and increases only linearly in time. In their scenario, the separation of neighboring field lines, which increases exponentially in time, plays a dominant role in triggering the onset of fast reconnection. 

The Boozer--Elder model is a welcome new development of Boozer's theory because it provides concrete and testable predictions. The primary objective of this study is to test some of the predictions. Specifically, we will focus on two distinct predictions, one related to ideal evolution and the other to resistive evolution. For the ideal evolution, the model predicts that the current density will increase linearly in time while the separation of neighboring field lines will grow exponentially in time. With a small but finite resistivity, because the exponential field-line separation amplifies the field line velocity, thereby speeding up reconnection, the \replace{}{time scale for the onset of fast reconnection and therefore the} current density will scale logarithmically with the Lundquist number.\footnote{A.~H.~Boozer, private communication (2022).} This logarithmic scaling relation for the current density is perhaps the most striking difference of Boozer's theory compared to traditional reconnection theories.

This paper is organized as follows. Section \ref{sec:Reduced-Magnetohydrodynamics-Mod} outlines the reduced magnetohydrodynamic (RMHD) model and the imposed footpoint motions. Section \ref{sec:Quasi-Static-Evolution} lays out the governing equations for the ideal and resistive quasi-static evolution of a coronal loop driven by footpoint motions. In Section \ref{sec:Ideal-Evolution}, we test Boozer and Elder's current density calculation with an ideal RMHD simulation. In Section \ref{sec:Resistive-Evolution}, we present resistive simulations to test the scaling of current density with the Lundquist number and investigate whether chaotic field line separation causes onset of reconnection. We conclude in Section \ref{sec:Conclusion}.  

\section{Reduced Magnetohydrodynamics Model \label{sec:Reduced-Magnetohydrodynamics-Mod}}

We employ the standard reduced magnetohydrodynamics (RMHD) model \citep{KadomtsevP1974,Strauss1976,VanBallegooijen1985} in this study. The RMHD model assumes a strong uniform guide field along the $z$ direction and that spatial scales along the guide field are much longer than that in perpendicular directions. Under these assumptions, the MHD equations can be simplified to a set of two equations
\begin{equation}
\partial_{t}\Omega+\text{\ensuremath{\left[\phi,\Omega\right]}}=\partial_{z}J+\left[A,J\right]+\nu\nabla_{\perp}^{2}\Omega-\lambda\Omega,\label{eq:RMHD-momentum}
\end{equation}
\begin{equation}
\partial_{t}A=\partial_{z}\phi+\left[A,\phi\right]+\eta\nabla_{\perp}^{2}A.\label{eq:RMHD-faraday}
\end{equation}
Here, we normalize the strength of the guide field to unity. The operator $\nabla_{\perp} \equiv \boldsymbol{\hat{x}}\partial_x + \boldsymbol{\hat{y}}\partial_y$ denotes the gradient operator in the perpendicular directions of the guide field. The magnetic
field is expressed in terms of the flux function $A$ through the
relation $\boldsymbol{B}=\boldsymbol{\hat{z}}+\nabla_{\perp}A\times\boldsymbol{\hat{z}}$.
The plasma velocity $\boldsymbol{u}$ is expressed in terms of the
stream function $\phi$ as $\boldsymbol{u}=\nabla_{\perp}\phi\times\boldsymbol{\hat{z}}$.
The vorticity and the electric current density along the $z$ direction
are given by $\Omega\equiv-\nabla_{\perp}^{2}\phi$ and $J\equiv-\nabla_{\perp}^{2}A$,
respectively. The Poisson bracket is defined as $\left[f,g\right]=\partial_{y}f\partial_{x}g-\partial_{x}f\partial_{y}g$.
Dissipation is introduced by including the resistivity $\eta$, the
viscosity $\nu$, and a friction coefficient $\lambda$. 

The RMHD model is widely used in analytic and numerical studies of Parker's coronal heating model \citep{VanBallegooijen1985,StraussO1988,LongcopeS1994a,LongcopeS1994b,NgB1998,DmitrukG1999,DmitrukGM2003,RappazzoVED2007,RappazzoVED2008,NgB2008,NgLB2012,RappazzoP2013}
and also in the study of Boozer and Elder.\citep{BoozerE2021} From the definitions of $\boldsymbol{B}$ and $\boldsymbol{u}$ and the Poisson bracket, we have the following useful relations 
\begin{equation}
\left[\phi,f\right]=\boldsymbol{u}\cdot\nabla f\label{eq:udg}
\end{equation}
and 
\begin{equation}
\partial_{z}f+\left[A,f\right]=\boldsymbol{B}\cdot\nabla f\label{eq:bdg}
\end{equation}
for an arbitrary variable $f$. 

The RMHD equations are solved with the DEBSRX code,\citep{HuangBB2014} which is a reduced version of the compressible MHD code DEBS.\citep{SchnackBMHCN1986}
The $x$--$y$ plane is discretized with a Fourier pseudospectral method. The $z$ direction is discretized with a finite-difference
scheme where $\phi$ and $A$ reside on staggered grids. The timestepping scheme is a semi-implicit, predictor-corrector leapfrog method where $\phi$ and $A$ are staggered in time. \replace{}{While the semi-implicit scheme allows time-steps to be larger than the limit set by the Courant--Friedrichs--Lewy (CFL) condition,\citep{CourantFL1928} too large a time-step may compromise accuracy. For this reason, we have been conservative in setting the time-step. We determine the time-step dynamically by using the CFL condition of the Alfv\'en wave along the guide field and that of the perpendicular flow speeds, whichever gives a smaller time-step. We have tested the accuracy of this choice by reducing the time steps by a factor of two for selected runs and have seen no significant difference. }

We assume that the system is bounded in the $z$ direction by two conducting plates at $z=0$ and $z=L$. The $x$--$y$ plane  is a $1\times1$ box ($0\le x,y\le1$) with doubly periodic boundary conditions. We take $L=10$ in this study. The bottom boundary at $z=0$ is stationary for all time; i.e., we impose the boundary condition $\phi=\phi_{b}=0$. At the top boundary of the simulation box ($z=L$), we impose a time-dependent flow given by the stream function
\begin{align}
\phi_{t}= & \left(\cos\left(2\pi x\right)-1\right)\left(\cos\left(2\pi y\right)-1\right)\nonumber \\
 & \left[c_{0}\sin\left(\omega_{0}t\right)+c_{1}\sin\left(2\pi x\right)\sin\left(\omega_{1}t\right)\right.\nonumber \\
 & \left.+c_{2}\sin\left(2\pi y\right)\sin\left(\omega_{2}t\right)\right.\nonumber \\
 & \left.+c_{3}\sin\left(2\pi x\right)\sin\left(2\pi y\right)\sin\left(\omega_{3}t\right)\right],\label{eq:boundary_phi}
\end{align}
where $c_{i}$ and $\omega_{i}$ are constants. This boundary flow
is similar, but not identical, to the flow employed by Boozer and
Elder.\citep{BoozerE2021} Because the DEBSRX code is limited to doubly-periodic
systems in the perpendicular directions whereas Boozer and Elder consider a cylindrical domain, we cannot impose the same boundary conditions
as they do. The difference in the boundary conditions, however, is
not germane to the issue we will discuss. Note that the boundary
flow is localized to the middle and vanishes at the edges of the simulation
box in the perpendicular directions. This feature is qualitatively
similar to the flow employed by Boozer and Elder.

We start the simulations with only the guide field. The imposed boundary flow drags the footpoints and gradually entangles the field lines. From solar observations, it is known that the time scales of footpoint motions are much longer compared with the Alfv\'en transit time along the coronal loop. We set the constants to $c_{0}=0$, $c_{1}=c_{2}=c_{3}=0.00025$,
$\omega_{1}=0.0125\sqrt{2}$, $\omega_{2}=0.0125\sqrt{3}$, and $\omega_{3}=0.0125\sqrt{5}$.
For this set of parameters, the maximal footpoint speed is typically of the order of $0.01$, corresponding to an advection time scale on the order of $100$. In contrast, the Alfv\'en transit time along the coronal
loop is $10$. Therefore, the condition for the separation of time
scales is met. We also make the ratios between frequencies $\omega_{i}/\omega_{j}$
irrational so that the flow pattern will not repeat itself, but this
choice is not crucial for the purpose of this study.

\section{Quasi-Static  Evolution of Coronal Loops\label{sec:Quasi-Static-Evolution}}

Under the condition that the characteristic time scales of footpoint
motions are much longer than the Alfv\'en transit time, the coronal
loop will evolve quasi-statically, provided that the coronal loop
is not close to an instability threshold. For a quasi-static coronal
loop, we may assume that the plasma inertia as well as the viscous
and frictional force are negligible, therefore the magnetic force-free
condition (within the RMHD approximation) 
\begin{equation}
\boldsymbol{B}\cdot\nabla J=\partial_{z}J+\left[A,J\right]=0\label{eq:force-balance}
\end{equation}
is satisfied for all time. The governing equations for quasi-static
evolution now are the force-free condition (\ref{eq:force-balance})
together with the induction equation (\ref{eq:RMHD-faraday}). In
this set of equations, the time derivative only appears in the induction
equation (\ref{eq:RMHD-faraday}), whereas the force-free condition
plays a role analogous to the incompressibility constraint, $\nabla\cdot\boldsymbol{u}=0$.

The quasi-static evolution of coronal loops driven by slow footpoint
motions can be determined as follows. First, taking the time derivative $\partial_{t}$
of Eq.~(\ref{eq:force-balance}) yields 
\begin{equation}
\boldsymbol{B}\cdot\nabla\partial_{t}J+\left[\partial_{t}A,J\right]=0.\label{eq:dt_force_balance}
\end{equation}
We can obtain an equation for $\partial_{t}J$ by applying the $-\nabla_{\perp}^{2}$
operator on Eq.~(\ref{eq:RMHD-faraday}), yielding
\begin{equation}
\partial_{t}J=-\nabla_{\perp}^{2}(\boldsymbol{B}\cdot\nabla\phi)+\eta\nabla_{\perp}^{2}J.\label{eq:dtJ1}
\end{equation}
Now, we can use Eq.~(\ref{eq:RMHD-faraday}) and Eq.~(\ref{eq:dtJ1}) to
eliminate the time derivatives in Eq.~(\ref{eq:dt_force_balance}) and obtain the following equation for $\phi$ 
\begin{equation}
\mathcal{L}\phi=\eta\boldsymbol{B}\cdot\nabla\left(\nabla_{\perp}^{2}J\right),\label{eq:qs3}
\end{equation}
where the linear operator $\mathcal{L}$ is defined as
\begin{equation}
\mathcal{L}\phi\equiv\boldsymbol{B}\cdot\nabla\left(\nabla_{\perp}^{2}(\boldsymbol{B}\cdot\nabla\phi)\right)-\left[\boldsymbol{B}\cdot\nabla\phi,J\right].\label{eq:L}
\end{equation}
At a given instant, if the operator $\mathcal{L}$ is invertible subject to the boundary conditions $\phi|_{z=0}=\phi_{b}$ and $\phi|_{z=L}=\phi_{t}$, we can in principle obtain the stream function $\phi$ (and the flow
$\boldsymbol{u}=\nabla_{\perp}\phi\times\boldsymbol{\hat{z}}$) that will carry the system to another force-free equilibrium at the next instant. In the ideal limit $\eta\to0$, Eq.~(\ref{eq:qs3}) is identical to the equation derived by van Ballegooijen in his study of the Parker problem.\citep{VanBallegooijen1985,NgB1998}

The operator $\mathcal{L}$ is the one that appears, unsurprisingly, also in the ideal linear stability problem of the instantaneous equilibrium, which takes the form
\begin{equation}
\mathcal{L}\phi=\gamma^{2}\nabla_{\perp}^{2}\phi.\label{eq:linear_stability}
\end{equation}
Here, a $\phi\sim e^{\gamma t}$ time-dependence and homogeneous boundary
conditions $\phi|_{z=0}=0$ and $\phi|_{z=L}=0$ are assumed. The
operators $\mathcal{L}$ and $\nabla_{\perp}^{2}$ in Eq.~(\ref{eq:linear_stability})
are self-adjoint; therefore, the eigenvalues $\gamma^{2}$ are real
and the eigenfunctions form a complete set; functions $\phi_{m}$
and $\phi_{n}$ of different eigenvalues satisfy the orthogonal condition
\begin{equation}
\int\nabla_{\perp}\phi_{m}^{*}\cdot\nabla_{\perp}\phi_{n}d^{3}x=0,\label{eq:orthogonal}
\end{equation}
while those with the same eigenvalues can be orthogonalized as well.\citep{ArfkenWH2013}

Suppose the complete set of eigenfunctions and eigenvalues $\{\phi_{n},\gamma_{n}^{2}\}$
is known. We can formally solve Eq.~(\ref{eq:qs3}) by making an eigenfunction expansion \footnote{The eigenfunction expansion should include the contribution of continuous spectra if they exist. For continuous spectra, the summation in Eq.~(\ref{eq:eigen_expansion}) should be replaced by an integral. However, continuous spectra are often associated with toroidal magnetic fields with nested flux surfaces. They are likely to be absent in the coronal loop model of this study due to the line-tied boundary condition.   }

\begin{equation}
\phi=\sum_{n}a_{n}\phi_{n}+\tilde{\phi}.\label{eq:eigen_expansion}
\end{equation}
Here, $\tilde{\phi}$ is an arbitrary smooth function satisfying the
inhomogeneous boundary conditions $\phi|_{z=0}=\phi_{b}$ and $\phi|_{z=L}=\phi_{t}$
{[}e.g., $\tilde{\phi}=\phi_{b}+\left(\phi_{t}-\phi_{b}\right)z/L${]}.
Using the orthogonality
of eigenfunctions, we can determine the coefficients $a_{n}$ as 
\begin{equation}
a_{n}=-\frac{\int\phi_{n}^{*}\left(\eta\boldsymbol{B}\cdot\nabla\left(\nabla_{\perp}^{2}J\right)-\mathcal{L}\tilde{\phi}\right)d^{3}x}{\gamma_{n}^{2}\int\nabla\phi_{n}^{*}\cdot\nabla\phi_{n}d^{3}x}.\label{eq:an}
\end{equation}
From Eq. (\ref{eq:an}), we conclude that Eq.~(\ref{eq:qs3}) is
solvable provided that the system is not at marginal stability, i.e.,
none of the eigenvalues $\gamma_{n}^{2}$ vanish. 

\section{Quasi-Static Ideal  Evolution\label{sec:Ideal-Evolution}}

\begin{figure}
\includegraphics[width=1\columnwidth]{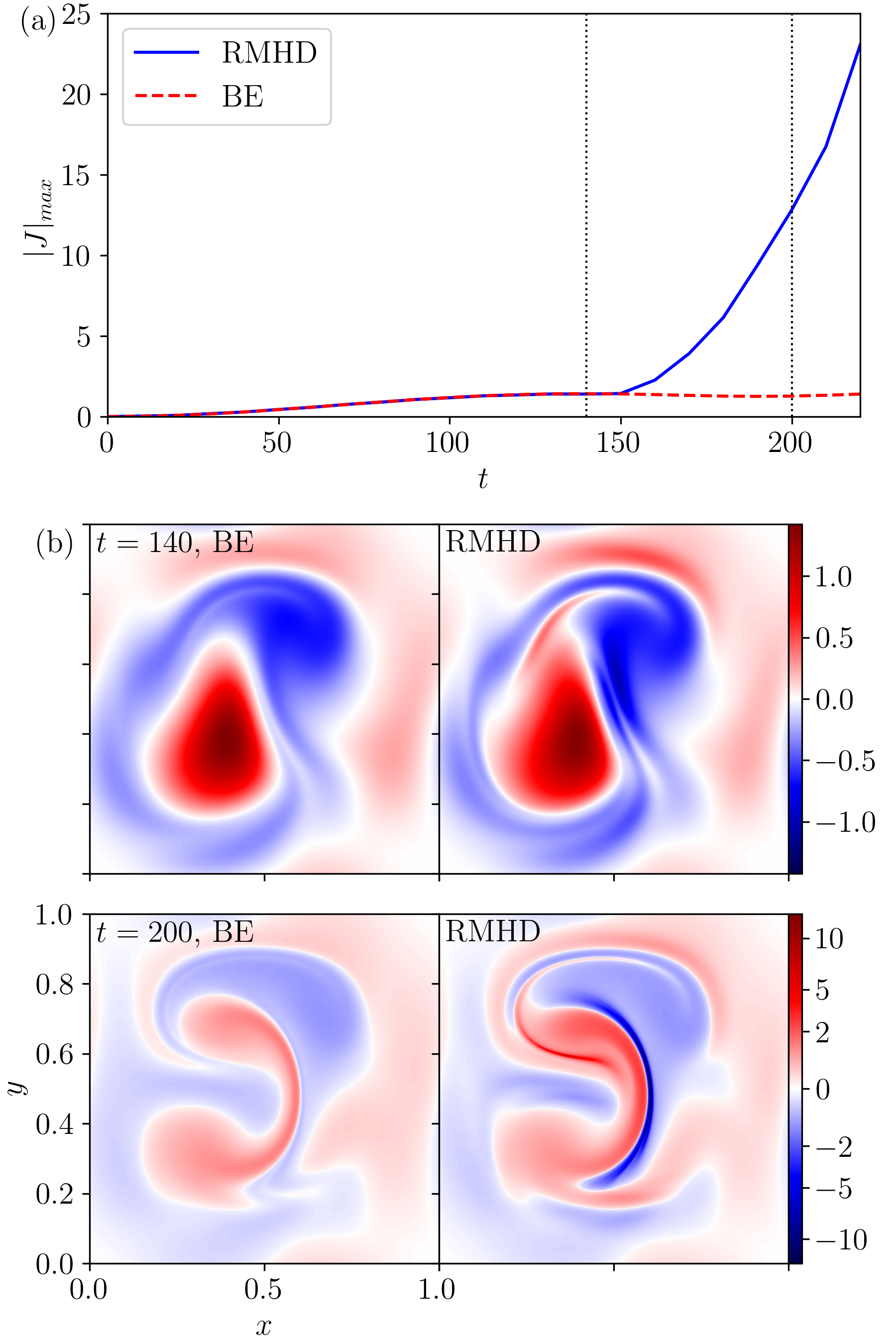}

\caption{Comparison between the current density at the top plate resulting from solving the BE equation and the RMHD equations. Panel (a) shows the time histories of the maximum current density obtained from both sets of solutions. The maximum current densities from both solutions agree up to $t=150$ and then depart substantially afterward. The current density in the RMHD solution increases significantly faster than predicted by the BE equation. Panel (b) shows snapshots from both sets of solutions at two representative times, at $t=140$ and $200$. Due to the significant difference in the current density range between the two solutions at $t=200$, the colorbar is stretched near $J=0$ for better visualization of the Boozer--Elder solution.
\label{fig:jz-comparison}}
\end{figure}

The above discussion shows that to determine the quasi-static evolution of this model
requires solving Eq.~(\ref{eq:qs3}) in the whole domain in each
time step. On the other hand, Boozer and Elder\citep{BoozerE2021}
take a different approach for the quasi-static ideal evolution that only involves solving an equation for the current density at the top boundary.
Their approach is as follows. First, using Eqs.~(\ref{eq:udg}), (\ref{eq:bdg}),
and the definition of the Poisson bracket, the term $-\nabla_{\perp}^{2}(\boldsymbol{B}\cdot\nabla\phi)$
in Eq.~(\ref{eq:dtJ1}) can be written as 
\begin{align}
-\nabla_{\perp}^{2}(\boldsymbol{B}\cdot\nabla\phi)= & -\partial_{z}\nabla_{\perp}^{2}\phi+\left[A,-\nabla_{\perp}^{2}\phi\right]+\left[-\nabla_{\perp}^{2}A,\phi\right]\nonumber \\
 & -\partial_{y}\nabla_{\perp}A\cdot\partial_{x}\nabla_{\perp}\phi+\partial_{x}\nabla_{\perp}A\cdot\partial_{y}\nabla_{\perp}\phi\nonumber \\
= & \boldsymbol{B}\cdot\nabla\Omega-\boldsymbol{u}\cdot\nabla J\nonumber \\
 & -\partial_{y}\boldsymbol{B}_{\perp}\cdot\partial_{x}\boldsymbol{u}+\partial_{x}\boldsymbol{B}_{\perp}\cdot\partial_{y}\boldsymbol{u}.\label{eq:del2_Bdphi}
\end{align}
Boozer and Elder ignore the terms $-\partial_{y}\boldsymbol{B}_{\perp}\cdot\partial_{x}\boldsymbol{u}+\partial_{x}\boldsymbol{B}_{\perp}\cdot\partial_{y}\boldsymbol{u}$,
whereupon Eq.~(\ref{eq:dtJ1}) becomes (here, we set $\eta=0$ for the ideal
evolution)
\begin{equation}
\partial_{t}J+\boldsymbol{u}\cdot\nabla J=\boldsymbol{B}\cdot\nabla\Omega.\label{eq:BE1}
\end{equation}
Because of the force-free condition $\boldsymbol{B}\cdot\nabla J=0$ and the ideal frozen-in condition, the left-hand-side of Eq.~(\ref{eq:BE1}) is constant along a field line. Consequently, the right-hand-side $\boldsymbol{B}\cdot\nabla\Omega$ is also constant along a field line. For the Boozer--Elder model, the vorticity at the bottom boundary vanishes and the vorticity at the top boundary is prescribed. Therefore, along a field line, $\boldsymbol{B}\cdot\nabla\Omega=\Omega|_{z=L}/L$ at any instant. Boozer and Elder then consider Eq.~(\ref{eq:BE1}) at $z=L$, yielding
\begin{equation}
\left(\partial_{t}J+\boldsymbol{u}\cdot\nabla J\right)_{z=L}=\frac{\Omega|_{z=L}}{L}.\label{eq:BE}
\end{equation}
This equation (hereafter the BE equation) plays a critical role in
the study of Boozer and Elder.\footnote{The derivation of the BE equation given by Boozer and Elder is slightly
different from ours. They start by writing Eq.~(\ref{eq:RMHD-faraday})(with
$\eta=0$) in the Lagrangian coordinates as $\left(\partial_{t}A\right)_{L}=\left(\partial_{z}\phi\right)_{L}$.
Here, the subscript $L$ implies the use of Lagrangian coordinates.
Next, they apply $-\nabla_{\perp}^{2}$ (also expressed in Lagrangian
coordinates) on both sides and obtain $\left(\partial_{t}J\right)_{L}=\left(\partial_{z}\Omega\right)_{L}$,
which is equivalent to Eq.~(\ref{eq:BE1}). However, this derivation
neglects the fact that the Laplacian $\nabla_{\perp}^{2}$ has explicit
time-dependence when it is expressed in Lagrangian coordinates because
the Lagrangian coordinates themselves are time-dependent. As such,
the Laplacian $\nabla_{\perp}^{2}$ and the time derivative $\partial_{t}$
do not commute. Consequently, $\left(\partial_{t}J\right)_{L}=\left(-\partial_{t}\nabla_{\perp}^{2}A\right)\neq\left(-\nabla_{\perp}^{2}\partial_{t}A\right)_{L}$,
implying some terms are missing in their derivation. This same issue
of missing terms also appears in some other publications, e.g., Eqs.~(D4)
and (D5) of Ref. {[}\onlinecite{Boozer2018}{]} and Eq.~(52) of Ref.
{[}\onlinecite{Boozer2022}{]}.} 

The BE equation takes the form of an advection equation for the current
density $J$ on the left-hand-side, whereas the right-hand-side serves
as a source term. Because the flow velocity $\boldsymbol{u}$ (and
therefore the vorticity $\Omega$) is prescribed at the top boundary,
the BE equation can be solved to determine the current density distribution at the top boundary without further knowledge of what occurs in the remaining part of the system. \replace{}{That is not the case if the missing terms are included, because $\boldsymbol{B}\cdot\nabla \Omega$ will  no longer be a constant along a field line. In other words, it is not possible to amend the BE equation simply by including those terms. }

\replace{}{The advection term on the left-hand-side of the BE equation only redistributes the current density without changing its magnitude. The peak current density can increase only through the source term on the right-hand-side. Hence, the current density is bounded by $\left| J \right|\le \Omega_{\text{max}}t/L$, where $ \Omega_{\text{max}}$ is an upper bound of $\left|\Omega\right|$ at the top boundary. In other words, the current density increases at most linearly in time according to the BE equation.}

Because some terms are neglected when deriving the BE equation, the question now is: are those terms negligible? We address this question by comparing solutions of the BE equation and that of the RMHD equations. The BE equation is solved
using a pseudospectral method implemented with the Dedalus framework;\citep{BurnsVOLB2020}
the grid resolution is $1024^{2}$. The RMHD equations are solved with the DEBSRX code with a grid resolution $1024^{3}$. We set $\eta=0$ in the DEBSRX simulation, but a small viscosity $\nu=10^{-6}$ is applied for numerical stability.  \replace{}{The numerical algorithms of the Dedalus implementation for the BE equation and the DEBSRX code for the RMHD equations are similar. Both use a Fourier pseudospectral method dealiased by the Orszag two-thirds rule\citep{Boyd2001} in the $x$--$y$ plane, so the numerical errors should be similar.} Both simulations have been compared with numerical solutions at lower resolutions \replace{}{($256^2$ and $512^2$ for BE; $256^3$ and $512^3$ for RMHD)} to ensure that the results presented here are well-resolved and converged.

Even though the time scales of footpoint motions are longer than the Alfv\'en transit time by approximately one order of magnitude, the RMHD calculation is not exactly force-free. To ensure that the RMHD solution remains close to force-free, we restart from a few selected snapshots, set the plasma speed to zero at footpoints and across the entire domain, then turn on the friction force and let the system relax to a force-free equilibrium. This experiment shows that the RMHD solution remains approximately force-free up to $t=200$, thereby ensuring a fair comparison between the BE and the RMHD solutions.

We now compare the RMHD solution at the top boundary with the BE solution and summarize the results in Figure \ref{fig:jz-comparison}. Panel (a) shows the time histories of the maximum current density obtained from both sets of solutions. The maximum current densities from both solutions agree until $t=150$ and then depart significantly afterward. Furthermore, the current density in the RMHD solution increases significantly faster than predicted by the BE equation. Panel (b) shows snapshots from both sets of solutions at two representative times, at $t=140$ and $200$. Although the two solutions give essentially the same maximum current density at $t=140$, we can already see differences between the two solutions. The differences become quite pronounced at $t=200$. By that time, thin current sheets have developed in the RMHD solution but are absent in the BE solution. 

Our results show that the neglected terms in the derivation of the BE equation are not negligible. Therefore, to determine the current density at the top boundary, we must solve for it over the entire 3D domain. Importantly, the BE equation significantly under-predicts the current density of the RMHD solution. As we will see in the next section, these intensifying current sheets eventually lead to the onset of reconnection. 

\section{Resistive Evolution\label{sec:Resistive-Evolution}}

\begin{table*}[t]
\begin{centering}
\begin{tabular}{ccccccccccccc}
\toprule 
Run & $S$ & $P_m$ & Resolution & $W_{\eta}$ & $W_{\nu}$ & $W_{\eta}+W_{\nu}$ & $W_{P}$ & $E_{M}$ & $E_{K}$ & $J_{1/2}$ & $V_{1/2}/V$\tabularnewline
\midrule 
A & $10^{4}$ & $1$ & $512^{3}$ & $1.36\ 10^{-2}$ & $2.66\ 10^{-4}$ & $1.38\ 10^{-2}$ &  $1.46\ 10^{-2}$ & $7.96\ 10^{-4}$ & $7.74\ 10^{-6}$ & 0.272 & $5.13\%$\tabularnewline
B & $4\ 10^{4}$ & $1$ & $512^{3}$ & $1.24\ 10^{-2}$ & $9.37\ 10^{-5}$ & $1.25\ 10^{-2}$ & $1.38\ 10^{-2}$ & $1.32\ 10^{-3}$ & $7.70\ 10^{-6}$ & 0.531 & $4.81\%$\tabularnewline
C & $10^{5}$ & $1$ & $512^{3}$ & $1.18\ 10^{-2}$ & $3.07\ 10^{-4}$ & $1.21\ 10^{-2}$ & $1.43\ 10^{-2}$ & $2.12\ 10^{-3}$ & $7.57\ 10^{-6}$ & 0.803 & $3.61\%$\tabularnewline
D1 & $4\ 10^{5}$ & $1$ & $512^{3}$ & $1.13\ 10^{-2}$ & $1.22\ 10^{-3}$ & $1.25\ 10^{-2}$ & $1.64\ 10^{-2}$ & $3.73\ 10^{-3}$ & $8.03\ 10^{-6}$ & 2.74 & $0.618\%$\tabularnewline
D2 & $4\ 10^{5}$ & $1$ & $768^{3}$ & $1.14\ 10^{-2}$ & $1.21\ 10^{-3}$ & $1.26\ 10^{-2}$ & $1.64\ 10^{-2}$ & $3.66\ 10^{-3}$ & $8.39\ 10^{-6}$ & 2.91 & $0.552\%$\tabularnewline
D3 & $4\ 10^{5}$ & $1$ & $1024^{3}$ & $1.14\ 10^{-2}$ & $1.21\ 10^{-3}$ & $1.26\ 10^{-2}$ & $1.63\ 10^{-2}$ & $3.64\ 10^{-3}$ & $8.34\ 10^{-6}$ & 2.90 & $0.550\%$\tabularnewline
E & $10^{6}$ & $1$ & $1024^{3}$ & $1.13\ 10^{-2}$ & $2.13\ 10^{-3}$ & $1.34\ 10^{-2}$ & $1.79\ 10^{-2}$ & $4.37\ 10^{-3}$ & $8.97\ 10^{-6}$ & 9.69 & $0.163\%$\tabularnewline
\bottomrule
\end{tabular}
\par\end{centering}
\caption{Parameters and energy diagnostic results of simulations reported in this paper. Here, $W_{\eta}=\int \eta J^2 \, d^3x \, dt$ and $W_{\nu}=\int \nu \Omega^2 \, d^3x \, dt$ are the resistive and viscous dissipation during the whole period of each simulation, respectively.  The Poynting energy input through the top boundary is $W_{P}=\int \boldsymbol{B}_\perp \cdot \boldsymbol{u} \, d^2x \, dt$. The magnetic energy $E_{M}=\int B_\perp^2/2 \, d^3x$ and the kinetic energy $E_{K}=\int u^2/2 \, d^3x$ are evaluated at the end of each simulation.  The error of the energy conservation relation $W_P = W_\eta + W_\nu + E_K + E_M$ is within 1\% for all cases. The current density $J_{1/2}$ indicates that regions with $\left|J\right|>J_{1/2}$ contribute half of the resistive dissipation, and $V_{1/2}/V$ is the time averaged volumetric ratio between regions with $\left|J\right|>J_{1/2}$ and the entire domain.\label{tab:Parameters}}
\end{table*}

\begin{figure}
\includegraphics[width=1\columnwidth]{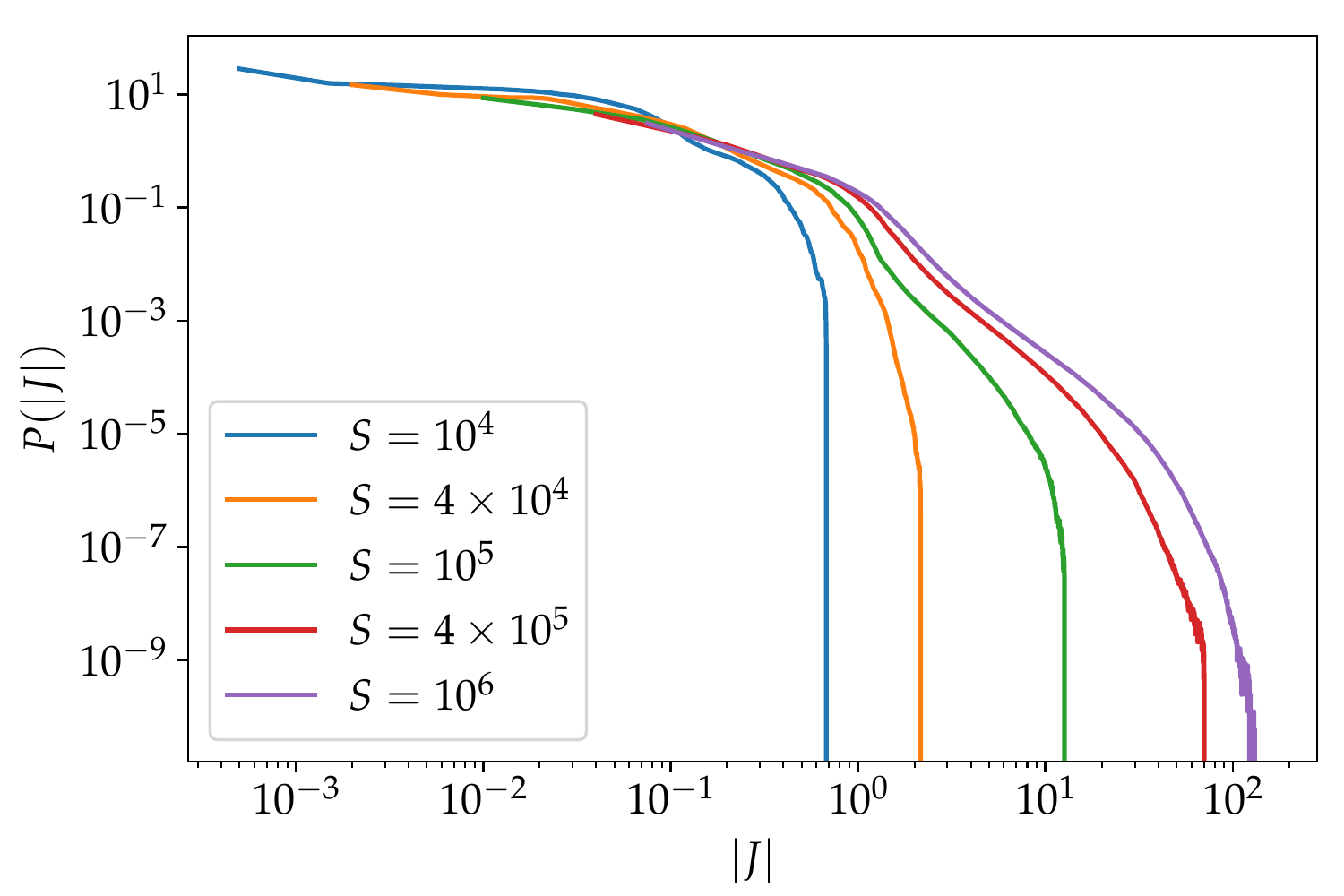}

\caption{Probability distributions of the scurrent density for various Lundquist
numbers $S$. Roughly speaking, the maximal current density scales linearly
with $S$. \label{fig:Probability-distribution-of-J}}
\end{figure}
\begin{figure}
\begin{centering}
\includegraphics[width=1\columnwidth]{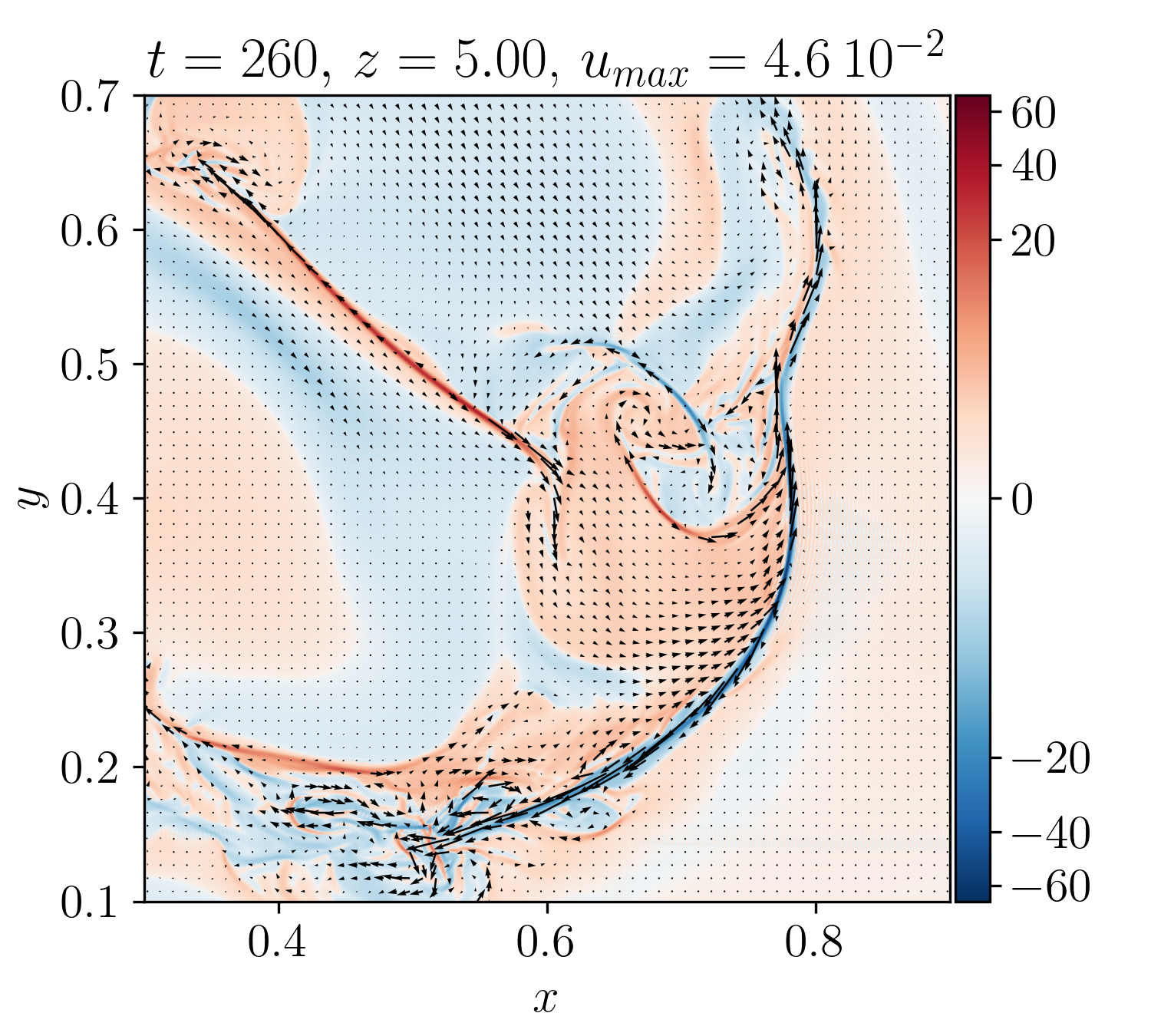}
\par\end{centering}
\caption{A 2D slice of Run E through the midplane $(z=5)$ at $t=260$. Here, we show the sub-domain of $0.3\le x\le0.9$ and $0.1\le y\le0.7$. The color shading shows the current density, and black arrows indicate
the plasma flow. Signatures of magnetic reconnection, such as outflow
jets within current sheets, are clearly visible. The entire time sequence
  is available as an animation (multimedia view). \label{fig:2D-slice}}
\end{figure}
\begin{figure*}
\centering{}\includegraphics[width=1.5\columnwidth]{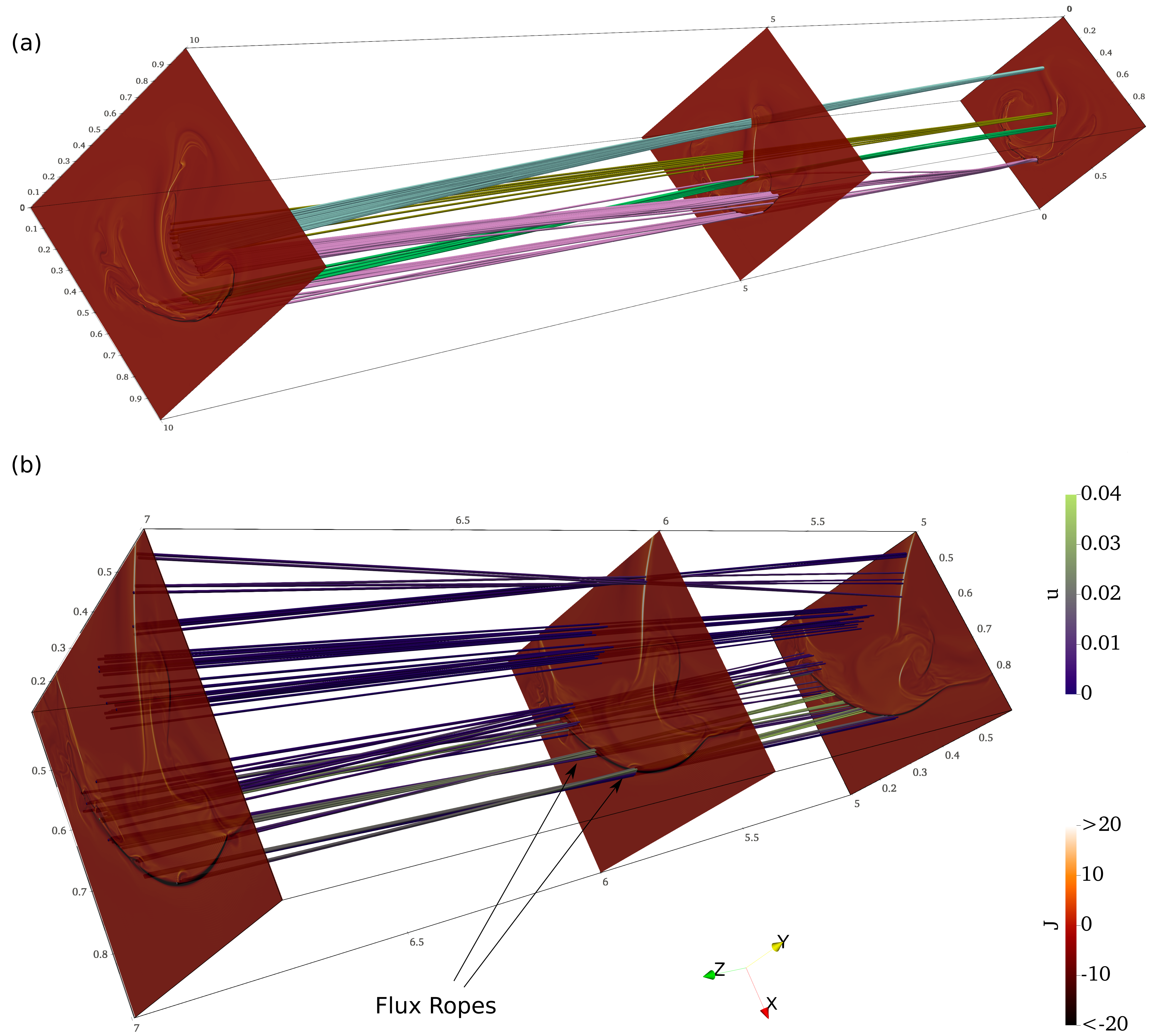}\caption{Three dimensional visualization of Run E at $t=260$. Panel (a) shows
the full domain. The color shading on the slices shows the current
density $J$ along the $z$ direction, indicating two primary current
ribbons with $J$ pointing at opposite directions. Panel (a) also
shows four bundles of field lines denoted with different colors. Each
of the bundle originates from a small area of radius 0.01 at the bottom boundary. As is typical for 3D magnetic fields, the field line bundles
spread out to much broader regions as we trace them along the $z$
direction. Panel (b) shows a zoom-in view of the sub-domain with $0.4\le x\le0.9$,
$0.1\le y\le0.6$, and $5\le z\le7$. Here, we see the current ribbon
with $J<0$ has become unstable the tearing (plasmoid) instability
and developed flux-rope-like structures within itself. The field lines in panel (b)
are color-coded according to the plasma flow speed $u$. \label{fig:Visualization-of-Run-E} }
\end{figure*}

We continue with the resistive evolution of the model problem. Our
objectives are to test the prediction that the peak current density
will scale logarithmically with respect to the Lundquist number $S$,
as well as to assess whether the ``exponential'' field line separation
causes the onset of reconnection.

To address these questions, we perform a series of simulations with
the Lundquist number $S$ varying from $10^{4}$ to $10^{6}$. Here,
the Lundquist number $S\equiv aV_{A}/\eta$ is defined through the
box size $a$ in the perpendicular direction, the Alfv\'en speed
$V_{A}$ of the guide field, and resistivity $\eta$. In our normalized
units, the box size $a=1$ and the Alfv\'en speed $V_{A}=1$; therefore
the Lundquist number is simply $S=1/\eta$. The viscosity is another
free parameter. For simplicity, we set the viscosity $\nu=\eta$ for
all cases; \replace{}{i.e., the magnetic Prandtl number $P_m\equiv \nu/\eta=1$. (See Appendix for a discussion of the effect of viscosity.) The friction coefficient $\lambda$ is set to zero.} Table \ref{tab:Parameters} lists the Lundquist numbers
and grid resolutions for all the simulations we have performed. The
simulation time is $t=1000$ for all cases, corresponding to 100 Alfv\'en
transit times \replace{}{and approximately ten footpoint advection times.}

Table \ref{tab:Parameters} also shows the total resistive dissipation
$W_{\eta}$ and viscous dissipation $W_{\nu}$ during the whole period
of each simulation. Over the range of Lundquist numbers $S$ that
spans two orders of magnitude, the total dissipation $W_{\eta}+W_{\nu}$
stays remarkably close to constant. The dissipation is predominantly
due to resistivity, although the portion of viscous dissipation slightly
increases as the Lundquist number increases. 

To ensure that our simulations have sufficient numerical resolution,
we perform three simulations (D1--D3) for the case $S=4\times10^{5}$
with resolutions ranging from $512^{3}$ to $1024^{3}$ and do not
find significant difference between them. Therefore, a grid resolution
of $512^{3}$ appears to be adequate for $S=4\times10^{5}$. This
finding gives us some confidence that the highest Lundquist number
case, Run E with $S=10^{6}$, should be reasonably resolved by a grid
resolution of $1024^{3}$. In addition, we may also assess the accuracy
of our numerical simulations by how precisely the energy is conserved.
The energy conservation requires that the total Poynting energy input through
the top boundary $W_{P}$ should be equal to the sum of the increase in magnetic energy $E_{M}$, the increase in kinetic energy $E_{K}$, and the total dissipation.
For all the cases, the error in energy conservation is less than 1\%
of the total Poynting energy input over the entire simulation period.

The column $J_{1/2}$ in Table \ref{tab:Parameters} indicates that regions
with $\left|J\right|>J_{1/2}$ contribute half of the resistive dissipation over the entire simulation period, and $V_{1/2}/V$ is the time averaged volumetric ratio between regions
with $\left|J\right|>J_{1/2}$ and the entire domain. These numbers
show that resistive dissipation is increasingly concentrated in small
regions of high current density as the Lundquist number increases.
For example, approximately $5\%$ of the total volume contributes
one half of the resistive dissipation at $S=10^{4}$, whereas less
than $0.2\%$ of the total volume accounts for the same portion at
$S=10^{6}$. 

We plot the probability distributions $P\left(\left|J\right|\right)$
of the current density \replace{}{over the four-dimensional spacetime during the period $0\le t\le1000$} for various Lundquist numbers $S$ in Figure
\ref{fig:Probability-distribution-of-J}. The probability distribution
exhibits a strong dependence on the Lundquist number. Furthermore,
the maximal current density roughly scales linearly with $S$. This
$J\sim S$ scaling relation is stronger than the Sweet--Parker\citep{Sweet1958a,Parker1957}
scaling relation $J\sim S^{1/2}$ and is on par with that of plasmoid-mediated
reconnection.\citep{HuangB2010} Therefore, the prediction that current
density will depend logarithmically on $S$ appears to be inconsistent
with our numerical findings. 

Now we address the question whether the exponential separation of
neighboring field lines causes onset of fast reconnection by taking
a closer look at the highest-$S$ simulation, Run E. 

During the early phase of the simulation when $t\le190$, the system evolves quasi-statically \replace{}{while the current density gradually increases as in the ideal evolution.} After $t=190$, \replace{}{which is approximately twice the footpoint advection time}, an onset of activity leads
to fast plasma flow \replace{}{and further intensifies the thin current sheets.} The peak
plasma flow speed after the onset is comparable to the typical in-plane
Alfv\'en speed and is faster than the typical footpoint speed by
an order of magnitude. In contrast, the plasma speed in the interior
is typically slower than or comparable to the typical footpoint speed
during the quasi-static phase. 

The flow pattern and current density distribution after the onset
clearly show signatures of magnetic reconnection, such as outflow
jets within current sheets, as can be seen in the 2D slice at $t=260$
shown in Figure \ref{fig:2D-slice}. Figure \ref{fig:Visualization-of-Run-E}
shows a 3D visualization of the entire domain as well as a zoom-in
view of a sub-domain. From the zoom-in view in panel (b), we can see
that the primary current sheet with $J<0$ (the one with dark colors)
has become unstable to the plasmoid (or tearing) instability and developed flux-rope-like structures despite the stabilizing effects of the line-tied boundary conditions. The plasmoid instability has continued to be present in numerous reconnecting current sheets throughout this simulation. Similar to systems without line-tied boundary conditions, the Lundquist number needs to exceed a threshold for the plasmoid instability to occur.\citep{BhattacharjeeHYR2009,HuangB2010,HuangCB2017,HuangCB2019} In all other simulations with lower Lundquist numbers, the plasmoid instability has not been observed. 

\replace{}{Previous studies indicate that the stabilizing effect of the line-tied boundary condition for the tearing mode is negligible when the ``geometric'' width $\delta_{\text{geo}}\sim (\lambda_{\text{tearing}}B_z/L B_{\text{up}})\delta$ is thinner than the inner layer width of the tearing mode, $\delta_{\text{tearing}}$.\citep{DelzannoF2008,HuangZ2009,RichardsonF2012, FinnBDZ2014} Here, $\delta$ is the current sheet thickness, $B_{\text{up}}$ is the upstream in-plane magnetic field of the current sheet, and $\lambda_{\text{tearing}}$ is the wavelength of the tearing mode. Taking the current sheet at $t=250$, immediately prior to the onset of the plasmoid instability, we estimate $B_{\text{up}} \simeq 0.1$ and $\delta\simeq 0.005$. For this set of parameters, the fastest growing mode wavelength $\lambda_{\text{tearing}} \simeq 0.15$ and the corresponding inner layer width $\delta_{\text{tearing}}\simeq0.001$,\citep{HuangCB2017} whereas the geometric width $\delta_{\text{geo}}\simeq 0.00075$. Therefore, the geometric width is comparable to the inner layer width, and the stabilizing effect of line-tying should be marginal. This estimate is consistent with the fact that the current sheet becomes unstable. In contrast, for cases of lower Lundquist numbers, the corresponding current sheets at the same time do not satisfy the criteria  $\delta_{\text{geo}}<\delta_{\text{tearing}}$ and remain stable.    } 

\begin{figure*}
\begin{centering}
\includegraphics[width=1\textwidth]{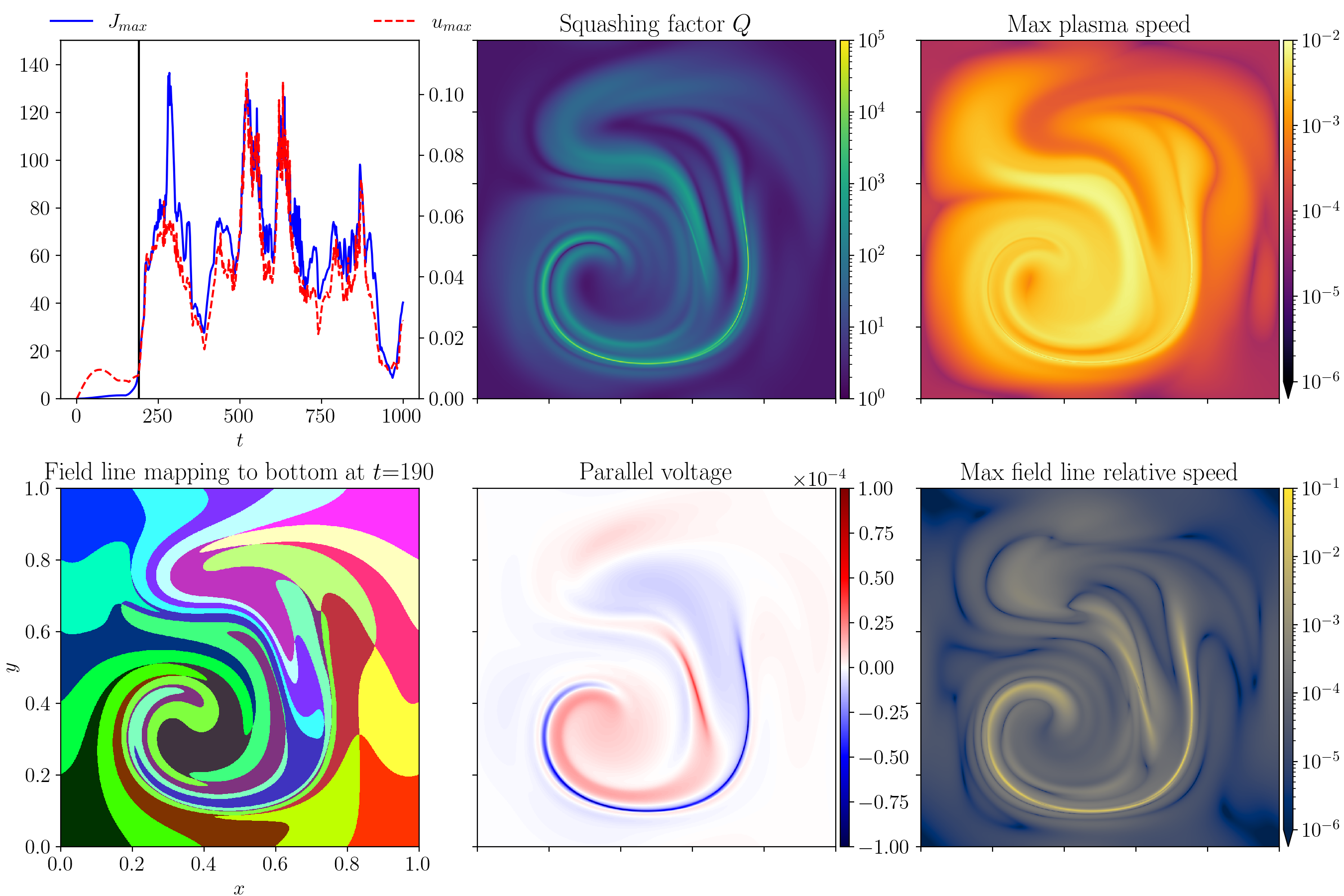}
\par\end{centering}
\caption{Diagnostic results at $t=190$. The upper-left panel shows the time
histories of the maximum current density $J_{\text{max}}$ and the
maximum plasma speed $u_{\text{max}}$ in the entire domain, with the vertical line indicating
the time of the snapshot. We label the field lines by their footpoints
at the bottom boundary. The lower-left panel shows the projected image
of $5\times5$ squares at the top boundary to the bottom boundary
following the field lines. The remaining four panels show the squashing
factor $Q$, the parallel voltage, the maximum plasma speed, and the
maximum field line relative speed along each field line.\label{fig:time1}}
\end{figure*}
\begin{figure*}
\begin{centering}
\includegraphics[width=1\textwidth]{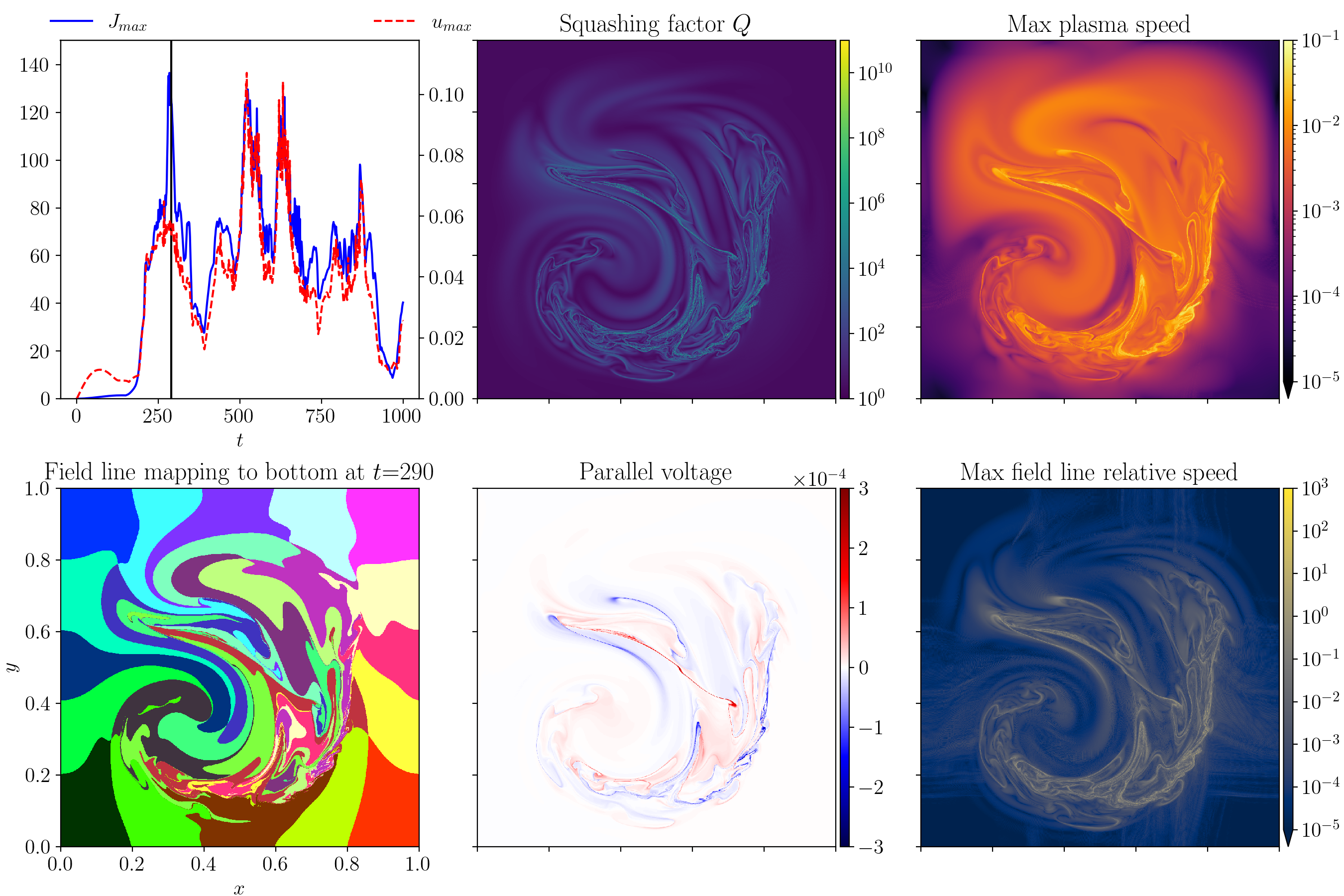}
\par\end{centering}
\caption{Diagnostic results at $t=290$. \label{fig:time2}}
\end{figure*}
\begin{figure*}
\begin{centering}
\includegraphics[width=1\textwidth]{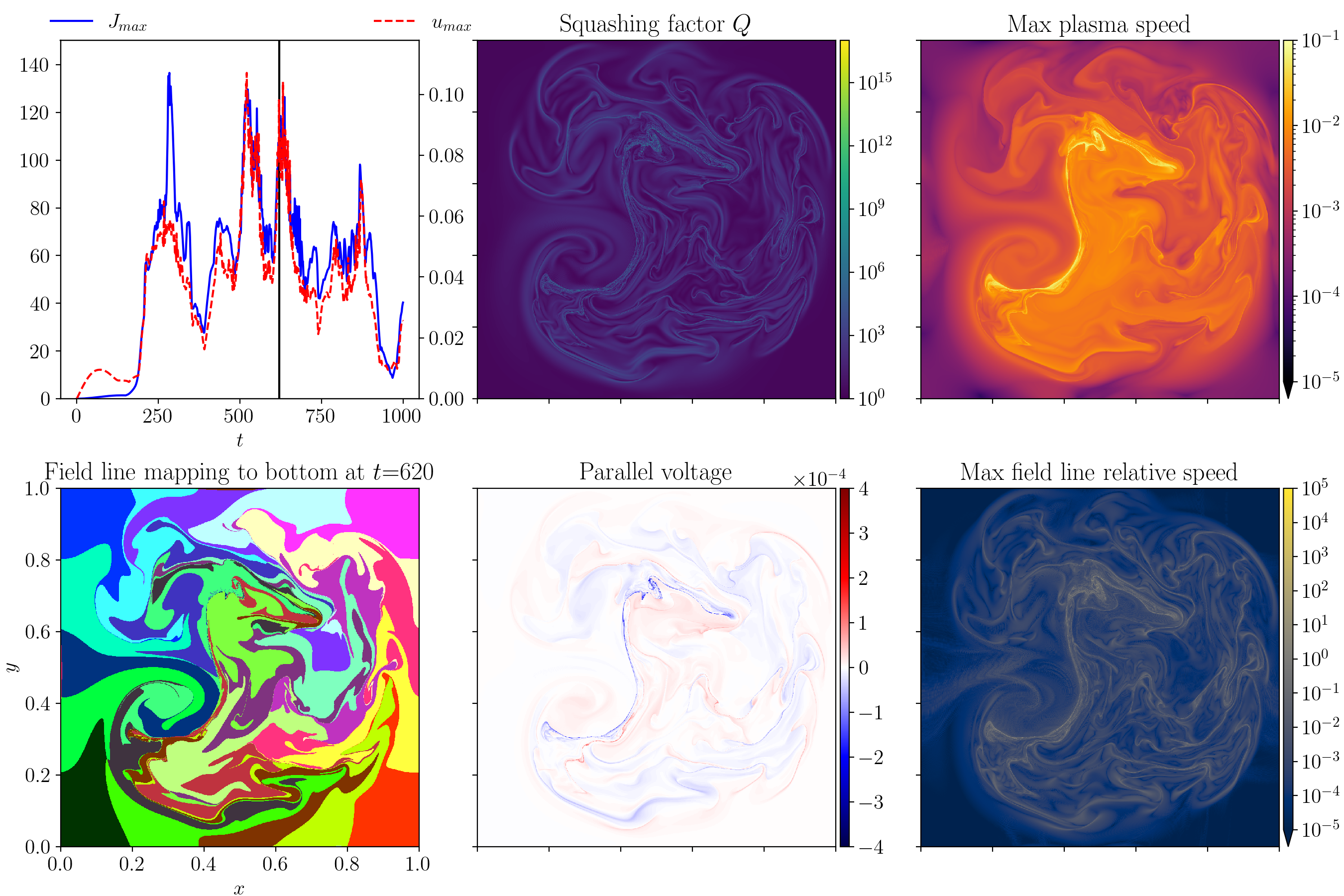}
\par\end{centering}
\caption{Diagnostic results at $t=620$. The entire time sequence of this Run
   is available as an animation (multimedia view). \label{fig:time3}}
\end{figure*}

To disentangle the relationship between reconnection and the exponential
separation of neighboring field lines, we have implemented a suite
of diagnostics together with field line tracing. Along each field
line, we calculate (a) the squashing factor $Q$,\citep{TitovH2002,Titov2007,FinnBDZ2014}
(b) the parallel voltage, (c) the plasma flow velocity, and (d) the
velocity of the field line in relative to the plasma. The squashing
factor $Q$ quantifies the extent of neighboring field line separation;
the parallel voltage is often employed as a metric for 3D reconnection
rate;\citep{SchindlerHB1988,SchindlerHB1991} finally, a large separation
between the field line velocity and the plasma velocity may also indicate
magnetic reconnection.

We calculate the squashing factor $Q$ by simultaneously integrating
the equation for magnetic field line flow,

\begin{equation}
\frac{d\boldsymbol{x}_{\perp}}{dz}=\boldsymbol{B}_{\perp},\label{eq:field_line}
\end{equation}
 and the equation for an infinitesimal separation $\delta\boldsymbol{x}_{\perp}$
between neighboring field lines 
\begin{equation}
\frac{d\delta\boldsymbol{x}_{\perp}}{dz}=\delta\boldsymbol{x}_{\perp}\cdot\nabla_{\perp}\boldsymbol{B}_{\perp}.\label{eq:separation}
\end{equation}
Equation (\ref{eq:separation}) can be written in matrix form as 
\begin{equation}
\frac{d}{dz}\left[\begin{array}{c}
\delta x\\
\delta y
\end{array}\right]=\left[\begin{array}{cc}
\partial_{x}B_{x} & \partial_{y}B_{x}\\
\partial_{x}B_{y} & \partial_{y}B_{y}
\end{array}\right]\left[\begin{array}{c}
\delta x\\
\delta y
\end{array}\right]\equiv M(z)\left[\begin{array}{c}
\delta x\\
\delta y
\end{array}\right].\label{eq:matrix_form}
\end{equation}
Because Eq.~(\ref{eq:matrix_form}) is linear in $\delta\boldsymbol{x}$,
it is sufficient to integrate it with respect to two linearly independent
initial condition. For that purpose, we integrate the equation 
\begin{equation}
\frac{dN}{dz}=MN\label{eq:matrix_eq}
\end{equation}
with the initial condition 
\begin{equation}
N|_{z=0}=\left[\begin{array}{cc}
1 & 0\\
0 & 1
\end{array}\right].\label{eq:initial_cond}
\end{equation}
Then, for any initial separation $\delta\boldsymbol{x}|_{z=0}$ we
can obtain the separation at an arbitrary $z$ as 
\begin{equation}
\left[\begin{array}{c}
\delta x\\
\delta y
\end{array}\right]=N(z)\left[\begin{array}{c}
\delta x\\
\delta y
\end{array}\right]_{z=0}.\label{eq:solution}
\end{equation}
The singular value decomposition (SVD) \citep{TrefethenB1997} of
$N(z)$ is of the form 
\begin{equation}
N(z)=U(z)\left[\begin{array}{cc}
\lambda_{\text{max}}(z) & 0\\
0 & \lambda_{\text{min}}(z)
\end{array}\right]V^{T}(z),\label{eq:SVD}
\end{equation}
where both $U(z)$ and $V(z)$ are unitary matrices. The singular
values $\lambda_{\text{max}}$ and $\lambda_{\text{min}}$ have a geometrical interpretation as follows. If we follow a infinitesimally
thin flux tube starting with a circular cross section of radius $\delta r$
at $z=0$, the cross section becomes an ellipse at $z>0$, with the
semi-major axis $\delta r_{\text{max}}=\lambda_{\text{max}}\delta r$
and the semi-minor-axis $\delta r_{\text{min}}=\lambda_{\text{min}}\delta r$.
Because the field line mapping in RMHD preserves area, the singular
values $\lambda_{\text{max}}$ and $\lambda_{\text{min}}$ satisfy
the relation $\lambda_{\text{max}}\lambda_{\text{min}}=1$. The squashing
factor $Q$ is then defined as
\begin{equation}
Q=\frac{\text{\ensuremath{\lambda}}_{\text{max}}}{\text{\ensuremath{\lambda}}_{\text{min}}}+\frac{\text{\ensuremath{\lambda}}_{\text{min}}}{\text{\ensuremath{\lambda}}_{\text{max}}}\label{eq:squashing}
\end{equation}
evaluated at the top plate. In the limit $Q\gg1$, the squashing factor
is approximately the ratio between the simi-major and the semi-minor
axes; i.e., $Q\simeq\text{\ensuremath{\lambda}}_{\text{max}}/\text{\ensuremath{\lambda}}_{\text{min}}$.
\footnote{Boozer and Elder employ the Frobenius norm of $N$, defined as $\left\Vert N\right\Vert =\sqrt{N_{xx}^{2}+N_{xy}^{2}+N_{yx}^{2}+N_{yy}^{2}}$,
to characterize the neighboring field line separation. The Frobenius
norm and the squashing factor are related by $\left\Vert N\right\Vert =\sqrt{Q}$.
Therefore, we can calculate $Q$ without invoking SVD.} 

Next, we calculate the field line velocity as follows. The electric
field in our RMHD model is given by the resistive Ohm's law
\begin{equation}
\boldsymbol{E}=-\boldsymbol{u}\times\boldsymbol{B}+\eta J\boldsymbol{\hat{z}}.\label{eq:Ohm}
\end{equation}
If we can express the electric field by 
\begin{equation}
\boldsymbol{E}=-\boldsymbol{v}\times\boldsymbol{B}+\nabla\Phi\label{eq:ideal}
\end{equation}
for some velocity field $\boldsymbol{v}$ and a scalar potential $\Phi$,
then the evolution of $\boldsymbol{B}$ is formally governed by the
ideal equation $\partial_{t}\boldsymbol{B}=\nabla\times\left(\boldsymbol{v}\times\boldsymbol{B}\right)$
such that the magnetic field lines are frozen-in to the velocity field
$\boldsymbol{v}$. The velocity field $\boldsymbol{v}$ is not uniquely
determined because we can add to $\boldsymbol{v}$ an arbitrary component
parallel to the magnetic field $\boldsymbol{B}$ without changing
Eq.~(\ref{eq:ideal}). Moreover, taking the inner product of Eq.~(\ref{eq:ideal})
and $\boldsymbol{B}$ yields 
\begin{equation}
\boldsymbol{B}\cdot\nabla\Phi=\eta J,\label{eq:BdPhi}
\end{equation}
which can be integrated along magnetic field lines to determine the
potential $\Phi$ up to a free function $\Phi(\boldsymbol{x}_{\perp})|_{z=0}$
defined on the bottom boundary. To uniquely specify $\boldsymbol{v}$,
we impose the conditions $v_{z}=0$ and $\Phi(\boldsymbol{x}_{\perp})|_{z=0}=0$.
The field line velocity $\boldsymbol{v}_{\perp}$ is then determined
by the relation
\begin{equation}
-\boldsymbol{\hat{z}}\times\boldsymbol{E}=\boldsymbol{u}_{\perp}=\boldsymbol{v}_{\perp}-\boldsymbol{\hat{z}}\times\nabla_{\perp}\Phi,\label{eq:v1}
\end{equation}
and the relative velocity between the field line and the plasma is given by
\begin{equation}
\boldsymbol{w}_{\perp}=\boldsymbol{v}_{\perp}-\boldsymbol{u}_{\perp}=\boldsymbol{\hat{z}}\times\nabla_{\perp}\Phi.\label{eq:w}
\end{equation}
The imposed boundary condition $\Phi(\boldsymbol{x}_{\perp})|_{z=0}=0$
makes the plasma velocity and the field line velocity coincide at
the bottom boundary. If the two velocities deviate significantly from
each other as we follow the field lines, it may indicate
that reconnection is ongoing. 

To calculate the field line velocity together with field-line tracing,
we adopt a method similar to the calculation of $Q$. Applying the
$\nabla_{\perp}$ operator on Eq.~(\ref{eq:BdPhi}) yields 
\begin{equation}
\partial_{z}\nabla_{\perp}\Phi+\boldsymbol{B}_{\perp}\cdot\nabla_{\perp}\nabla_{\perp}\Phi=-\nabla_{\perp}\boldsymbol{B}_{\perp}\cdot\nabla_{\perp}\Phi+\eta\nabla_{\perp}J.\label{eq:Bdggphi}
\end{equation}
The left-hand-side of Eq.~(\ref{eq:Bdggphi}) corresponds to the
variation of $\nabla_{\perp}\Psi$ along field lines. Using Eq.~(\ref{eq:w})
to replace $\nabla_{\perp}\Psi$ by $\boldsymbol{w}_{\perp}$, we
can rewrite Eq.~(\ref{eq:Bdggphi}) in a matrix form 
\begin{equation}
\frac{d}{dz}\left[\begin{array}{c}
w_{x}\\
w_{y}
\end{array}\right]=\left[\begin{array}{cc}
-\partial_{y}B_{y} & \partial_{y}B_{x}\\
\partial_{x}B_{y} & -\partial_{x}B_{x}
\end{array}\right]\left[\begin{array}{c}
w_{x}\\
w_{y}
\end{array}\right]+\eta\left[\begin{array}{c}
-\partial_{y}J\\
\partial_{x}J
\end{array}\right].\label{eq:dzw}
\end{equation}
Here, the derivative $d/dz$ on the left-hand-side is a derivative
along the magnetic field lines. We can now integrate Eq.~(\ref{eq:dzw})
together with field line tracing to obtain the field line velocity. 

Figures \ref{fig:time1}, \ref{fig:time2}, and \ref{fig:time3} show
the results of the diagnostics at three representative times. In each
figure, the upper-left panel shows the time histories of the maximum
current density $J_{\text{max}}$ and the maximum plasma speed $u_{\text{max}}$,
with the vertical line indicating the time of the snapshot. We label
the field lines by their footpoints at the bottom boundary. The lower-left
panel shows the projected image of $5\times5$ squares at the top
boundary to the bottom boundary following the field lines. The remaining
four panels show the squashing factor $Q$, the parallel voltage $\int_{0}^{L}\eta J\,dz$,
the maximum plasma speed $\left|\boldsymbol{u}\right|$, and the maximum
field line relative speed $\left|\boldsymbol{w}\right|$ along each
field line. At each time slice, we trace $1000\times1000$ field lines
uniformly distributed at the bottom boundary. Additionally, we provide
the entire time history of these diagnostics for Run E, together with
that for Runs A, B, C, and D3 as animations (see multimedia view in Fig.~\ref{fig:time3} and Supplementary Material).

Figure \ref{fig:time1} shows that the maximum $Q$ is approaching
$10^{5}$ at $t=190$. The ``slippage'' speed between field line
and the plasma also become approximately one order of magnitude larger
than the plasma speed for field lines with high $Q$ values. The maximum
slippage of footpoint mapping relative to that of the ideal run is
approximately $30\%$ of the simulation box size in the perpendicular
direction. Although one might interpret that significant magnetic
reconnection has already occurred based on this information, it may be
more appropriate to attribute the footpoint slippage to diffusion
rather than reconnection because the evolution remains close to quasi-static
up to this time.

After $t=190$, there is a rapid onset of activity with the current
density and plasma speed both increasing substantially. \replace{}{Notably, the onset occurs in approximately two footpoint advection times, much sooner than in ten advection times predicted by Boozer and Elder.} Subsequently, at $t=290$
shown in Fig.~\ref{fig:time2}, the plasma speed is approximately
one order of magnitude higher than that during the quasi-static phase,
the squashing factor $Q$ goes above $10^{10}$, and the field line
relative speed is above $10^{2}$, which is more than three orders
of magnitude higher than the plasma speed. The correlations between
the squashing factor $Q$, the parallel voltage, and the plasma relative
speed are also evident. The same general features are also present
in Fig.~\ref{fig:time3} for $t=620$. At this time, the squashing
factor $Q$ goes beyond $10^{15}$ at some locations, and the maximal
field line relative speed is above $10^{4}$, more than five orders
of magnitude higher than the plasma speed. 

Based on the result that both the squashing factor and the field line
speed become extremely high, would one conclude that chaotic field
line separation causes fast reconnection? Upon close examination,
our simulation results do not appear to support this viewpoint. If
chaotic field-line separation is the cause of fast reconnection, we
expect the squashing factor to reach a maximum when the coronal loop
``breaks loose,'' i.e., when it starts to deviate from quasi-static
evolution; subsequently, reconnection will simplify the field-line
mapping and the squashing factor will decrease. That is not what the
simulation shows. As we can see from the time sequence shown in Figures
\ref{fig:time1} -- \ref{fig:time3} and the associated animation,
after the coronal loop ``breaks loose'' after $t=190$, thin current
sheets intensify and the squashing factor increases tremendously around
them. Unlike Boozer's claim that chaotic field-line separation can
speed up reconnection without intense current sheets, the simulation
shows that current sheets must intensify to release the built-up magnetic
stress; the intensified current sheets further enhance chaotic field-line separation as reflected in a even higher squashing factor $Q$.\footnote{Note that even though the existence of intense current sheets is not
a necessary condition for a high squashing factor $Q$, the former
naturally leads to the later because of the strongly sheared magnetic
field associated with thin current sheets.} 

For a complicated 3D evolution with reconnection occurring at multiple
locations and at different times, it is difficult, if not impossible,
to quantify the speed of the reconnection process with a single reconnection
rate. However, we may quantify the effectiveness of reconnection through
its effect of converting magnetic energy into plasma heating through
dissipation. Because the dissipation rate is nearly independent of
the Lundquist number $S$, it is not unreasonable to think that reconnection
proceed approximately at the same rate for different $S$. This conclusion
is consistent with the fact that the peak values of the parallel voltage,
often employed as a metric of 3D reconnection rate, mostly fluctuate
between $2\times10^{-4}$ and $6\times10^{-4}$, regardless
of the Lundquist number. In contrast, the field-line speed exhibits
a strong dependence on the Lundquist number $S$. At $S=10^{4}$,
the field line relative speed stays below $10^{-2}$; whereas at $S=10^{6}$
, the plasma relative speed can go above $10^{4}$. Our findings suggest
that the parallel voltage is a more reliable indicator of the reconnection
rate than the field-line speed.

\section{Discussion and Conclusions\label{sec:Conclusion}}

In conclusion, we have tested Boozer's reconnection theory using the Boozer--Elder model for a coronal loop driven by footpoint motions. Our simulation results significantly differ from their predictions in both ideal and resistive evolution.

For the ideal evolution, we show that Boozer and Elder significantly under-predict the intensity of electric current in the coronal loop due to missing terms in their equations. For the resistive evolution, our simulations show that the maximal current density roughly scales linearly with the Lundquist number $S$, in stark contrast to the prediction of a logarithmic dependence on $S$. \replace{}{Because of the formation of intense thin current sheets, the onset of fast reconnection occurs much sooner than predicted by Boozer and Elder.} Therefore, our simulation results do not appear to support Boozer's theory. \replace{}{Moreover, thin current sheets become unstable to the plasmoid instability when the Lundquist number is sufficiently high.}

\replace{}{A precise definition of 3D reconnection remains an open question. Boozer's definition of reconnection relies entirely on the connections between fluid elements, and he attributes any changes in the connections to reconnection. This definition, while precise, is overly general, and blurs the distinction between reconnection and diffusion. As an illuminating example, let us first consider a stable line-tied screw pinch\citep{ZweibelB1985, HuangZS2006} undergoing resistive diffusion. Because the footpoint mapping between the boundaries changes as time progresses, this process is reconnection according to Boozer's definition, although it occurs on a slow, resistive time scale. Next, let us consider the same process in a stable coronal loop with chaotic field lines. The time evolution of the magnetic field remains slow, but the field-line velocity is fast due to the amplification caused by the chaotic field-line mapping. This enhanced field-line speed is fast reconnection according to Boozer's definition. Examples of slow resistive diffusion with rapid changes in field-line connections have been demonstrated in Ref.~[\onlinecite{HuangBB2014}]. However, the same study also shows that when the slow resistive diffusion leads to an unstable configuration, the time evolution becomes dynamic, intense thin current sheets form, and reconnection outflow jets ensue. At that time, field-line connections change even faster.}

\replace{}{These results of Ref.~[\onlinecite{HuangBB2014}]} and the present study indicate that field-line velocity is not a reliable metric for reconnection rate. We should keep in mind that field-line velocity is a mathematical construct rather than a physical velocity. Furthermore, the field-line speed has no upper limit and can even exceed the speed of light! Therefore, we should be cautious about using field line velocity to draw conclusions. In comparison, the parallel voltage, which is the integration of the parallel electric field, appears to be a better indicator of the reconnection rate. Because the parallel voltage $\Phi\sim\eta JL$ and the maximal current
density $J_{\text{max}}\propto S\propto1/\eta$, the maximal voltage
$\Phi_{\text{max}}$ is approximately independent of $S$. Therefore,
the reconnection rate in the coronal loop is approximately independent
of $S$. This conclusion is consistent with the dissipation rate being
insensitive to $S$.

Our simulation results \replace{}{show that intense thin current sheets are a natural outcome of a coronal loop forced by slow footpoint motions.} This conclusion appears to be more consistent with Parker's
nanoflare scenario, where thin current sheets play a crucial role,
than with Boozer's theory. However, Boozer's theory does provide some
new perspectives and poses new challenges to Parker's scenario. In
particular, Boozer points out the fragility of the ideal MHD frozen-in
constraint in the presence of chaotic field lines. This important
aspect warrants a broader attention and further investigation. Traditionally,
the Parker problem is usually posed as a question regarding whether
singular current sheets can form in a coronal loop under the
frozen-in constraint and line-tied boundary conditions. This viewpoint
attributes the formation of current sheets solely as an ideal MHD
effect. However, because the footpoint motion at the photosphere naturally
renders the field line chaotic, the ideal MHD frozen-in constraint
may be overly restrictive for the Parker problem. In this study, we
have observed that resistive simulations can develop more intensive current sheets than an ideal simulation at the same time when both are still in a quasi-static phase, indicating that non-ideal effects may play an active role in the formation of current sheets.\citep{HuangBB2014} A similar suggestion has been made by Bhattacharjee and Wang, who proposed that helicity-conserving reconnection processes might facilitate the formation of current sheets without rendering the footpoint velocity discontinuous.\citep{BhattacharjeeW1991} The roles of non-ideal effects in current sheet formation and onset of reconnection in coronal loops will be a topic of future study.

\section{Supplementary Material}

Diagnostic results for Runs A, B, C, and D3, similar to those shown in Fig.~\ref{fig:time3}, are available as animations.

\begin{acknowledgments}
This research was supported by the U.S. Department of Energy, grant
number DE-SC0021205, and the National Aeronautics and Space Administration,
grant numbers 80NSSC18K1285 and 88NSSC21K1326. Computations were performed
on facilities at National Energy Research Scientific Computing Center.
We thank Professor Allen Boozer for numerous beneficial discussion
about his view on 3D reconnection, as well as the anonymous referees for insightful comments. This paper is dedicated to the memory of Aad van Ballegooijen, who made incisive contributions to the problem of current sheets in coronal loops and whose untimely passing in 2021 has deprived the community of a gentleman and a scholar. 
\end{acknowledgments}

\section*{Data Availability}
The data that support the findings of this study are available from the corresponding author upon reasonable request.

\appendix*
\replace{}{\section*{Appendix: The Effects of Viscosity}}

\begin{table*}[t]
\begin{centering}
\begin{tabular}{ccccccccccccc}
\toprule 
Run & $S$ & $P_m$ & Resolution & $W_{\eta}$ & $W_{\nu}$ & $W_{\eta}+W_{\nu}$ & $W_{P}$ & $E_{M}$ & $E_{K}$ & $J_{1/2}$ & $V_{1/2}/V$\tabularnewline
\midrule 
F & $10^{4}$ & $10$ & $512^{3}$ & $1.37\ 10^{-2}$ & $2.51\ 10^{-3}$ & $1.62\ 10^{-2}$ & $1.70\ 10^{-2}$ & $7.96\ 10^{-4}$ & $7.77\ 10^{-6}$ & 0.273 & $5.10\%$\tabularnewline
G & $ 10^{5}$ & $10^2$ &$512^{3}$ & $1.28\ 10^{-2}$ & $3.65\ 10^{-3}$ & $1.65\ 10^{-2}$ & $1.85\ 10^{-2}$ & $2.04\ 10^{-3}$ & $7.72\ 10^{-6}$ & 0.885 & $3.52\%$\tabularnewline
H & $10^{6}$ & $10^3$ & $1024^{3}$ & $1.29\ 10^{-2}$ & $6.50\ 10^{-3}$ & $1.94\ 10^{-2}$ & $2.43\ 10^{-2}$ & $4.78\ 10^{-3}$ & $7.71\ 10^{-6}$ & 4.98 & $0.81\%$\tabularnewline

\bottomrule
\end{tabular}
\par\end{centering}
\caption{Parameters and energy diagnostic results of simulations reported in the Appendix. See the caption of Table \ref{tab:Parameters} for the definition of each column. These runs keep a constant value of the viscosity $\nu=10^{-3}$ and vary the resistivity $\eta$, whereas the runs in Table \ref{tab:Parameters} have $P_m=1$. \label{tab:Parameters1}}
\end{table*}

\begin{figure}
\includegraphics[width=1\columnwidth]{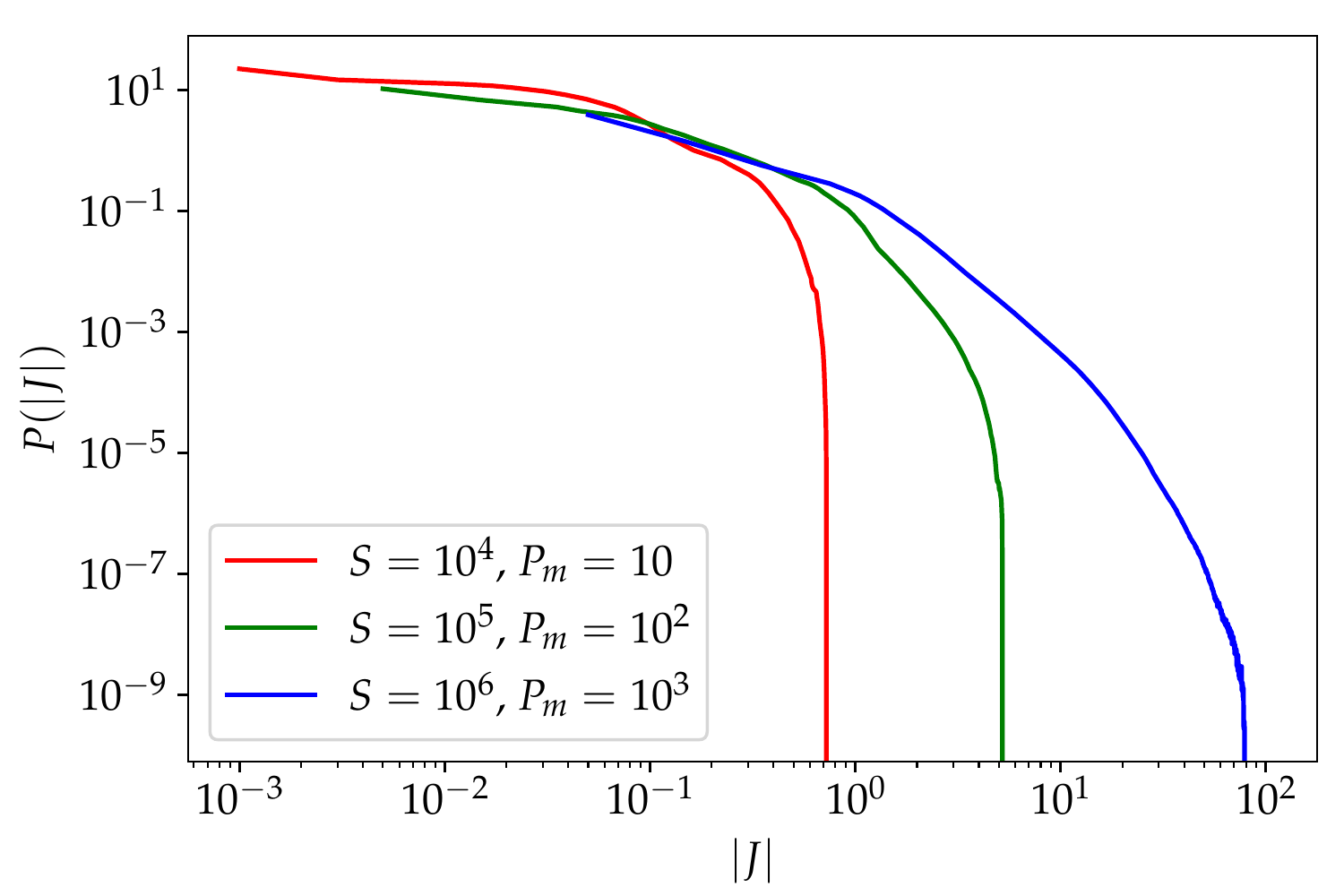}

\caption{Probability distributions of the current density for various Lundquist
  numbers $S$, with the viscosity fixed at $\nu=10^{-3}$.  \label{fig:Probability-distribution-of-J-visc}}
\end{figure}
\begin{figure}
\begin{centering}
\includegraphics[width=1\columnwidth]{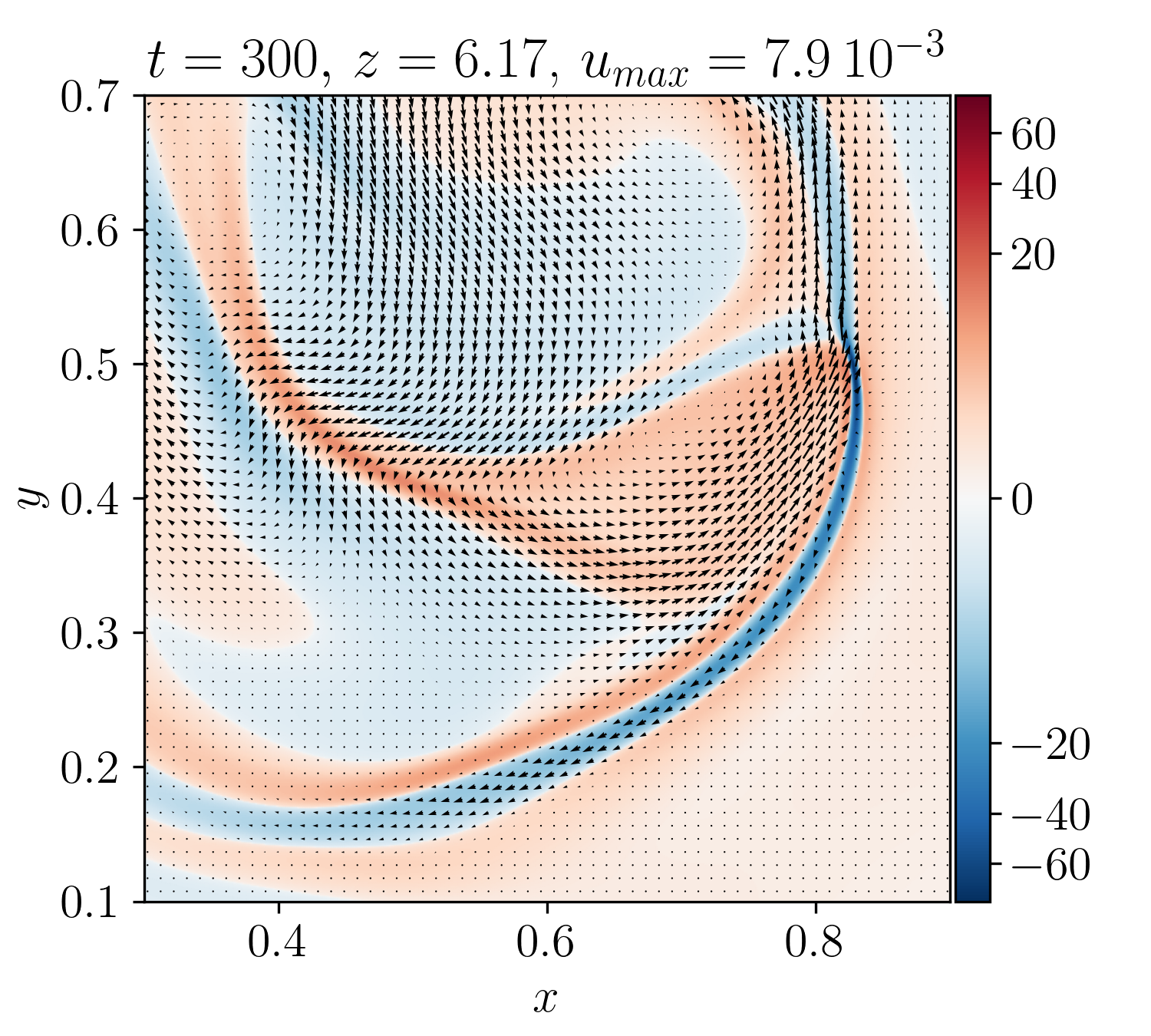}
\par\end{centering}
\caption{A 2D slice of Run H at $t=300$. Here, we show the sub-domain of $0.3\le x\le0.9$ and $0.1\le y\le0.7$. The color shading shows the current density, and black arrows indicate
the plasma flow. \label{fig:2D-slice-visc}}
\end{figure}

\replace{}{After the initial submission of this paper, Boozer suggested that the observed signatures of reconnection, in particular the intense thin current sheets, may be attributed to damping of Alfv\'en waves after changes of the field-line connections release the stored magnetic energy as plasma kinetic energy. He further suggested that by increasing the viscosity to damp the kinetic energy, intense thin current sheets may disappear.\citep{Boozer2022a} This hypothesis is interesting and has practical relevance as well, because observational evidence suggests that the magnetic Prandtl number $P_m \equiv \nu/\eta$ of coronal loops may be orders of magnitudes larger than unity.\citep{Aschwanden2005}}

\replace{}{We test this hypothesis by performing additional simulations with $P_m \gg 1$. We keep the viscosity at a constant value $\nu=10^{-3}$, which is as high as possible without significantly compromising the force-free approximation of the magnetic field as the coronal loop evolves quasi-statically. The values of $\eta$ vary from $10^{-4}$ to $10^{-6}$. }

\replace{}{Table \ref{tab:Parameters1} summarizes the parameters and diagnostic results of the high-$P_m$ runs. Remarkably, resistive dissipation still dominates over viscous dissipation even when $P_m\gg 1$. The resistive dissipation remains close to constant as $\eta$ varies. Moreover, compared with the results in Table \ref{tab:Parameters}, the resistive dissipation for $P_m\gg 1$ is similar to that for $P_m=1$.} 

\replace{}{Figure \ref{fig:Probability-distribution-of-J-visc} shows the probability distribution of the current density for various Lundquist numbers $S$. Similar to Figure \ref{fig:Probability-distribution-of-J} for $P_m=1$, here the probability distribution also exhibits a strong dependence on the Lundquist number $S$, although the maximal current density is lower than the corresponding maximal current density with the same $S$ and $P_m=1$. A 2D slice of Run H with $S=10^6$ and $P_m=10^3$ is shown in Fig.~\ref{fig:2D-slice-visc}. 
Compared with Fig.~\ref{fig:2D-slice}, the plasma velocity is significantly smoother and slower due to the higher viscosity, whereas the current density is only slightly smoother. }

\replace{}{Even though increasing the magnetic Prandtl number does smooth the current sheets, the maximal current density still exhibits a strong dependence on the Lundquist number $S$.  Therefore, intense thin current sheets appear to be a natural consequence of this coronal loop model with high Lundquist numbers, regardless of the magnetic Prandtl number.}
\bibliographystyle{apsrev4-2}
\bibliography{ref}

\begin{thebibliography}{88}%
\makeatletter
\providecommand \@ifxundefined [1]{%
 \@ifx{#1\undefined}
}%
\providecommand \@ifnum [1]{%
 \ifnum #1\expandafter \@firstoftwo
 \else \expandafter \@secondoftwo
 \fi
}%
\providecommand \@ifx [1]{%
 \ifx #1\expandafter \@firstoftwo
 \else \expandafter \@secondoftwo
 \fi
}%
\providecommand \natexlab [1]{#1}%
\providecommand \enquote  [1]{``#1''}%
\providecommand \bibnamefont  [1]{#1}%
\providecommand \bibfnamefont [1]{#1}%
\providecommand \citenamefont [1]{#1}%
\providecommand \href@noop [0]{\@secondoftwo}%
\providecommand \href [0]{\begingroup \@sanitize@url \@href}%
\providecommand \@href[1]{\@@startlink{#1}\@@href}%
\providecommand \@@href[1]{\endgroup#1\@@endlink}%
\providecommand \@sanitize@url [0]{\catcode `\\12\catcode `\$12\catcode
  `\&12\catcode `\#12\catcode `\^12\catcode `\_12\catcode `\%12\relax}%
\providecommand \@@startlink[1]{}%
\providecommand \@@endlink[0]{}%
\providecommand \url  [0]{\begingroup\@sanitize@url \@url }%
\providecommand \@url [1]{\endgroup\@href {#1}{\urlprefix }}%
\providecommand \urlprefix  [0]{URL }%
\providecommand \Eprint [0]{\href }%
\providecommand \doibase [0]{https://doi.org/}%
\providecommand \selectlanguage [0]{\@gobble}%
\providecommand \bibinfo  [0]{\@secondoftwo}%
\providecommand \bibfield  [0]{\@secondoftwo}%
\providecommand \translation [1]{[#1]}%
\providecommand \BibitemOpen [0]{}%
\providecommand \bibitemStop [0]{}%
\providecommand \bibitemNoStop [0]{.\EOS\space}%
\providecommand \EOS [0]{\spacefactor3000\relax}%
\providecommand \BibitemShut  [1]{\csname bibitem#1\endcsname}%
\let\auto@bib@innerbib\@empty
\bibitem [{\citenamefont {Boozer}(2012{\natexlab{a}})}]{Boozer2012}%
  \BibitemOpen
  \bibfield  {author} {\bibinfo {author} {\bibfnamefont {A.~H.}\ \bibnamefont
  {Boozer}},\ }\href {https://doi.org/10.1063/1.4765352} {\bibfield  {journal}
  {\bibinfo  {journal} {Phys. Plasmas}\ }\textbf {\bibinfo {volume} {19}},\
  \bibinfo {pages} {112901} (\bibinfo {year} {2012}{\natexlab{a}})}\BibitemShut
  {NoStop}%
\bibitem [{\citenamefont {Boozer}(2012{\natexlab{b}})}]{Boozer2012a}%
  \BibitemOpen
  \bibfield  {author} {\bibinfo {author} {\bibfnamefont {A.~H.}\ \bibnamefont
  {Boozer}},\ }\href {https://doi.org/10.1063/1.4754715} {\bibfield  {journal}
  {\bibinfo  {journal} {Phys. Plasmas}\ }\textbf {\bibinfo {volume} {19}},\
  \bibinfo {pages} {092902} (\bibinfo {year} {2012}{\natexlab{b}})}\BibitemShut
  {NoStop}%
\bibitem [{\citenamefont {Boozer}(2013)}]{Boozer2013}%
  \BibitemOpen
  \bibfield  {author} {\bibinfo {author} {\bibfnamefont {A.~H.}\ \bibnamefont
  {Boozer}},\ }\href {https://doi.org/10.1063/1.4796051} {\bibfield  {journal}
  {\bibinfo  {journal} {Phys. Plasmas}\ }\textbf {\bibinfo {volume} {20}},\
  \bibinfo {pages} {032903} (\bibinfo {year} {2013})}\BibitemShut {NoStop}%
\bibitem [{\citenamefont {Boozer}(2014)}]{Boozer2014}%
  \BibitemOpen
  \bibfield  {author} {\bibinfo {author} {\bibfnamefont {A.~H.}\ \bibnamefont
  {Boozer}},\ }\href {https://doi.org/10.1063/1.4890491} {\bibfield  {journal}
  {\bibinfo  {journal} {Phys. Plasmas}\ }\textbf {\bibinfo {volume} {21}},\
  \bibinfo {pages} {072907} (\bibinfo {year} {2014})}\BibitemShut {NoStop}%
\bibitem [{\citenamefont {Boozer}(2018)}]{Boozer2018}%
  \BibitemOpen
  \bibfield  {author} {\bibinfo {author} {\bibfnamefont {A.~H.}\ \bibnamefont
  {Boozer}},\ }\href {https://doi.org/10.1017/S0022377818000028} {\bibfield
  {journal} {\bibinfo  {journal} {Journal of Plasma Physics}\ }\textbf
  {\bibinfo {volume} {84}},\ \bibinfo {pages} {715840102} (\bibinfo {year}
  {2018})}\BibitemShut {NoStop}%
\bibitem [{\citenamefont {Boozer}(2019)}]{Boozer2019}%
  \BibitemOpen
  \bibfield  {author} {\bibinfo {author} {\bibfnamefont {A.~H.}\ \bibnamefont
  {Boozer}},\ }\href {https://doi.org/10.1063/1.5094179} {\bibfield  {journal}
  {\bibinfo  {journal} {Physics of Plasmas}\ }\textbf {\bibinfo {volume}
  {26}},\ \bibinfo {pages} {082112} (\bibinfo {year} {2019})}\BibitemShut
  {NoStop}%
\bibitem [{\citenamefont {Boozer}(2021)}]{Boozer2021}%
  \BibitemOpen
  \bibfield  {author} {\bibinfo {author} {\bibfnamefont {A.~H.}\ \bibnamefont
  {Boozer}},\ }\href {https://doi.org/10.1063/5.0031413} {\bibfield  {journal}
  {\bibinfo  {journal} {Physics of Plasmas}\ }\textbf {\bibinfo {volume}
  {28}},\ \bibinfo {pages} {032102} (\bibinfo {year} {2021})}\BibitemShut
  {NoStop}%
\bibitem [{\citenamefont {Boozer}(2022{\natexlab{a}})}]{Boozer2022}%
  \BibitemOpen
  \bibfield  {author} {\bibinfo {author} {\bibfnamefont {A.~H.}\ \bibnamefont
  {Boozer}},\ }\href {https://doi.org/10.1063/5.0089793} {\bibfield  {journal}
  {\bibinfo  {journal} {Physics of Plasmas}\ }\textbf {\bibinfo {volume}
  {29}},\ \bibinfo {pages} {052104} (\bibinfo {year}
  {2022}{\natexlab{a}})}\BibitemShut {NoStop}%
\bibitem [{\citenamefont {Boozer}\ and\ \citenamefont
  {Elder}(2021)}]{BoozerE2021}%
  \BibitemOpen
  \bibfield  {author} {\bibinfo {author} {\bibfnamefont {A.~H.}\ \bibnamefont
  {Boozer}}\ and\ \bibinfo {author} {\bibfnamefont {T.}~\bibnamefont {Elder}},\
  }\href {https://doi.org/10.1063/5.0039776} {\bibfield  {journal} {\bibinfo
  {journal} {Physics of Plasmas}\ }\textbf {\bibinfo {volume} {28}},\ \bibinfo
  {pages} {062303} (\bibinfo {year} {2021})}\BibitemShut {NoStop}%
\bibitem [{\citenamefont {Biskamp}(2000)}]{Biskamp2000}%
  \BibitemOpen
  \bibfield  {author} {\bibinfo {author} {\bibfnamefont {D.}~\bibnamefont
  {Biskamp}},\ }\href@noop {} {\emph {\bibinfo {title} {Magnetic Reconnection
  in Plasmas}}}\ (\bibinfo  {publisher} {Cambridge University Press},\ \bibinfo
  {year} {2000})\BibitemShut {NoStop}%
\bibitem [{\citenamefont {Priest}\ and\ \citenamefont
  {Forbes}(2000)}]{PriestF2000}%
  \BibitemOpen
  \bibfield  {author} {\bibinfo {author} {\bibfnamefont {E.~R.}\ \bibnamefont
  {Priest}}\ and\ \bibinfo {author} {\bibfnamefont {T.}~\bibnamefont
  {Forbes}},\ }\href@noop {} {\emph {\bibinfo {title} {Magnetic reconnection :
  {MHD} theory and applications}}}\ (\bibinfo  {publisher} {Cambridge
  University Press},\ \bibinfo {year} {2000})\BibitemShut {NoStop}%
\bibitem [{\citenamefont {Zweibel}\ and\ \citenamefont
  {Yamada}(2009)}]{ZweibelY2009}%
  \BibitemOpen
  \bibfield  {author} {\bibinfo {author} {\bibfnamefont {E.~G.}\ \bibnamefont
  {Zweibel}}\ and\ \bibinfo {author} {\bibfnamefont {M.}~\bibnamefont
  {Yamada}},\ }\href {https://doi.org/10.1146/annurev-astro-082708-101726}
  {\bibfield  {journal} {\bibinfo  {journal} {Annu. Rev. Astron. Astrophys.}\
  }\textbf {\bibinfo {volume} {47}},\ \bibinfo {pages} {291} (\bibinfo {year}
  {2009})}\BibitemShut {NoStop}%
\bibitem [{\citenamefont {Yamada}\ \emph {et~al.}(2010)\citenamefont {Yamada},
  \citenamefont {Kulsrud},\ and\ \citenamefont {Ji}}]{YamadaKJ2010}%
  \BibitemOpen
  \bibfield  {author} {\bibinfo {author} {\bibfnamefont {M.}~\bibnamefont
  {Yamada}}, \bibinfo {author} {\bibfnamefont {R.}~\bibnamefont {Kulsrud}},\
  and\ \bibinfo {author} {\bibfnamefont {H.}~\bibnamefont {Ji}},\ }\href
  {https://doi.org/10.1103/RevModPhys.82.603} {\bibfield  {journal} {\bibinfo
  {journal} {Rev. Mod. Phys.}\ }\textbf {\bibinfo {volume} {82}},\ \bibinfo
  {pages} {603} (\bibinfo {year} {2010})}\BibitemShut {NoStop}%
\bibitem [{\citenamefont {Zweibel}\ and\ \citenamefont
  {Yamada}(2016)}]{ZweibelY2016}%
  \BibitemOpen
  \bibfield  {author} {\bibinfo {author} {\bibfnamefont {E.~G.}\ \bibnamefont
  {Zweibel}}\ and\ \bibinfo {author} {\bibfnamefont {M.}~\bibnamefont
  {Yamada}},\ }\href {https://doi.org/10.1098/rspa.2016.0479} {\bibfield
  {journal} {\bibinfo  {journal} {Proc. R Soc. A}\ }\textbf {\bibinfo {volume}
  {472}},\ \bibinfo {pages} {20160479} (\bibinfo {year} {2016})}\BibitemShut
  {NoStop}%
\bibitem [{\citenamefont {Ji}\ \emph {et~al.}(2022)\citenamefont {Ji},
  \citenamefont {Daughton}, \citenamefont {Jara-Almonte}, \citenamefont {Le},
  \citenamefont {Stanier},\ and\ \citenamefont {Yoo}}]{JiDJLSY2022}%
  \BibitemOpen
  \bibfield  {author} {\bibinfo {author} {\bibfnamefont {H.}~\bibnamefont
  {Ji}}, \bibinfo {author} {\bibfnamefont {W.}~\bibnamefont {Daughton}},
  \bibinfo {author} {\bibfnamefont {J.}~\bibnamefont {Jara-Almonte}}, \bibinfo
  {author} {\bibfnamefont {A.}~\bibnamefont {Le}}, \bibinfo {author}
  {\bibfnamefont {A.}~\bibnamefont {Stanier}},\ and\ \bibinfo {author}
  {\bibfnamefont {J.}~\bibnamefont {Yoo}},\ }\href
  {https://doi.org/10.1038/s42254-021-00419-x} {\bibfield  {journal} {\bibinfo
  {journal} {Nature Reviews Physics}\ }\textbf {\bibinfo {volume} {4}},\
  \bibinfo {pages} {263} (\bibinfo {year} {2022})}\BibitemShut {NoStop}%
\bibitem [{\citenamefont {Pontin}\ and\ \citenamefont
  {Priest}(2022)}]{PontinP2022}%
  \BibitemOpen
  \bibfield  {author} {\bibinfo {author} {\bibfnamefont {D.~I.}\ \bibnamefont
  {Pontin}}\ and\ \bibinfo {author} {\bibfnamefont {E.~R.}\ \bibnamefont
  {Priest}},\ }\href {https://doi.org/10.1007/s41116-022-00032-9} {\bibfield
  {journal} {\bibinfo  {journal} {Living Reviews in Solar Physics}\ }\textbf
  {\bibinfo {volume} {19}},\ \bibinfo {pages} {1} (\bibinfo {year}
  {2022})}\BibitemShut {NoStop}%
\bibitem [{\citenamefont {Priest}\ \emph {et~al.}(2003)\citenamefont {Priest},
  \citenamefont {Hornig},\ and\ \citenamefont {Pontin}}]{PriestHP2003}%
  \BibitemOpen
  \bibfield  {author} {\bibinfo {author} {\bibfnamefont {E.~R.}\ \bibnamefont
  {Priest}}, \bibinfo {author} {\bibfnamefont {G.}~\bibnamefont {Hornig}},\
  and\ \bibinfo {author} {\bibfnamefont {D.~I.}\ \bibnamefont {Pontin}},\
  }\href {https://doi.org/10.1029/2002JA009812} {\bibfield  {journal} {\bibinfo
   {journal} {Journal of Geophysical Research}\ }\textbf {\bibinfo {volume}
  {108}},\ \bibinfo {pages} {1285} (\bibinfo {year} {2003})}\BibitemShut
  {NoStop}%
\bibitem [{\citenamefont {Pontin}(2011)}]{Pontin2011}%
  \BibitemOpen
  \bibfield  {author} {\bibinfo {author} {\bibfnamefont {D.~I.}\ \bibnamefont
  {Pontin}},\ }\href {https://doi.org/10.1016/j.asr.2010.12.022} {\bibfield
  {journal} {\bibinfo  {journal} {Adv. Space Res.}\ }\textbf {\bibinfo {volume}
  {47}},\ \bibinfo {pages} {1508} (\bibinfo {year} {2011})}\BibitemShut
  {NoStop}%
\bibitem [{\citenamefont {Greene}(1993)}]{Greene1993}%
  \BibitemOpen
  \bibfield  {author} {\bibinfo {author} {\bibfnamefont {J.~M.}\ \bibnamefont
  {Greene}},\ }\href@noop {} {\bibfield  {journal} {\bibinfo  {journal} {Phys.
  Fluids B}\ }\textbf {\bibinfo {volume} {5}},\ \bibinfo {pages} {2355}
  (\bibinfo {year} {1993})}\BibitemShut {NoStop}%
\bibitem [{\citenamefont {Schindler}\ \emph {et~al.}(1988)\citenamefont
  {Schindler}, \citenamefont {Hesse},\ and\ \citenamefont
  {Birn}}]{SchindlerHB1988}%
  \BibitemOpen
  \bibfield  {author} {\bibinfo {author} {\bibfnamefont {K.}~\bibnamefont
  {Schindler}}, \bibinfo {author} {\bibfnamefont {M.}~\bibnamefont {Hesse}},\
  and\ \bibinfo {author} {\bibfnamefont {J.}~\bibnamefont {Birn}},\ }\href@noop
  {} {\bibfield  {journal} {\bibinfo  {journal} {J. Geophy. Res.}\ }\textbf
  {\bibinfo {volume} {93}},\ \bibinfo {pages} {5547} (\bibinfo {year}
  {1988})}\BibitemShut {NoStop}%
\bibitem [{\citenamefont {Hesse}\ and\ \citenamefont
  {Schindler}(1988)}]{HesseS1988}%
  \BibitemOpen
  \bibfield  {author} {\bibinfo {author} {\bibfnamefont {M.}~\bibnamefont
  {Hesse}}\ and\ \bibinfo {author} {\bibfnamefont {K.}~\bibnamefont
  {Schindler}},\ }\href@noop {} {\bibfield  {journal} {\bibinfo  {journal} {J.
  Geophy. Res.}\ }\textbf {\bibinfo {volume} {93}},\ \bibinfo {pages} {5559}
  (\bibinfo {year} {1988})}\BibitemShut {NoStop}%
\bibitem [{\citenamefont {Schindler}\ \emph {et~al.}(1991)\citenamefont
  {Schindler}, \citenamefont {Hesse},\ and\ \citenamefont
  {Birn}}]{SchindlerHB1991}%
  \BibitemOpen
  \bibfield  {author} {\bibinfo {author} {\bibfnamefont {K.}~\bibnamefont
  {Schindler}}, \bibinfo {author} {\bibfnamefont {M.}~\bibnamefont {Hesse}},\
  and\ \bibinfo {author} {\bibfnamefont {J.}~\bibnamefont {Birn}},\ }\href@noop
  {} {\bibfield  {journal} {\bibinfo  {journal} {Astrophys. J.}\ }\textbf
  {\bibinfo {volume} {380}},\ \bibinfo {pages} {293} (\bibinfo {year}
  {1991})}\BibitemShut {NoStop}%
\bibitem [{\citenamefont {Titov}\ and\ \citenamefont
  {Hornig}(2002)}]{TitovH2002}%
  \BibitemOpen
  \bibfield  {author} {\bibinfo {author} {\bibfnamefont {V.~S.}\ \bibnamefont
  {Titov}}\ and\ \bibinfo {author} {\bibfnamefont {G.}~\bibnamefont {Hornig}},\
  }\href@noop {} {\bibfield  {journal} {\bibinfo  {journal} {Adv. Space Res.}\
  }\textbf {\bibinfo {volume} {29}},\ \bibinfo {pages} {1087} (\bibinfo {year}
  {2002})}\BibitemShut {NoStop}%
\bibitem [{\citenamefont {Titov}(2007)}]{Titov2007}%
  \BibitemOpen
  \bibfield  {author} {\bibinfo {author} {\bibfnamefont {V.~S.}\ \bibnamefont
  {Titov}},\ }\href@noop {} {\bibfield  {journal} {\bibinfo  {journal}
  {Astrophys. J.}\ }\textbf {\bibinfo {volume} {660}},\ \bibinfo {pages} {863}
  (\bibinfo {year} {2007})}\BibitemShut {NoStop}%
\bibitem [{\citenamefont {Titov}\ \emph {et~al.}(2009)\citenamefont {Titov},
  \citenamefont {Forbes}, \citenamefont {Priest}, \citenamefont {Miki\'c},\
  and\ \citenamefont {Linker}}]{TitovFPML2009}%
  \BibitemOpen
  \bibfield  {author} {\bibinfo {author} {\bibfnamefont {V.~S.}\ \bibnamefont
  {Titov}}, \bibinfo {author} {\bibfnamefont {T.~G.}\ \bibnamefont {Forbes}},
  \bibinfo {author} {\bibfnamefont {E.~R.}\ \bibnamefont {Priest}}, \bibinfo
  {author} {\bibfnamefont {Z.}~\bibnamefont {Miki\'c}},\ and\ \bibinfo {author}
  {\bibfnamefont {J.~A.}\ \bibnamefont {Linker}},\ }\href@noop {} {\bibfield
  {journal} {\bibinfo  {journal} {Astrophys. J.}\ }\textbf {\bibinfo {volume}
  {693}},\ \bibinfo {pages} {1029} (\bibinfo {year} {2009})}\BibitemShut
  {NoStop}%
\bibitem [{\citenamefont {Parker}(1972)}]{Parker1972}%
  \BibitemOpen
  \bibfield  {author} {\bibinfo {author} {\bibfnamefont {E.~N.}\ \bibnamefont
  {Parker}},\ }\href@noop {} {\bibfield  {journal} {\bibinfo  {journal}
  {Astrophys. J.}\ }\textbf {\bibinfo {volume} {174}},\ \bibinfo {pages} {499}
  (\bibinfo {year} {1972})}\BibitemShut {NoStop}%
\bibitem [{\citenamefont {Parker}(1988)}]{Parker1988}%
  \BibitemOpen
  \bibfield  {author} {\bibinfo {author} {\bibfnamefont {E.~N.}\ \bibnamefont
  {Parker}},\ }\href@noop {} {\bibfield  {journal} {\bibinfo  {journal}
  {Astrophys. J.}\ }\textbf {\bibinfo {volume} {330}},\ \bibinfo {pages} {474}
  (\bibinfo {year} {1988})}\BibitemShut {NoStop}%
\bibitem [{\citenamefont {Parker}(1994)}]{Parker1994}%
  \BibitemOpen
  \bibfield  {author} {\bibinfo {author} {\bibfnamefont {E.~N.}\ \bibnamefont
  {Parker}},\ }\href@noop {} {\emph {\bibinfo {title} {Spontaneous Current
  Sheets in Magnetic Fields}}}\ (\bibinfo  {publisher} {Oxford University
  Press, Inc.},\ \bibinfo {year} {1994})\BibitemShut {NoStop}%
\bibitem [{\citenamefont {van Ballegooijen}(1985)}]{VanBallegooijen1985}%
  \BibitemOpen
  \bibfield  {author} {\bibinfo {author} {\bibfnamefont {A.~A.}\ \bibnamefont
  {van Ballegooijen}},\ }\href@noop {} {\bibfield  {journal} {\bibinfo
  {journal} {Astrophys. J.}\ }\textbf {\bibinfo {volume} {298}},\ \bibinfo
  {pages} {421} (\bibinfo {year} {1985})}\BibitemShut {NoStop}%
\bibitem [{\citenamefont {Zweibel}\ and\ \citenamefont
  {Li}(1987)}]{ZweibelL1987}%
  \BibitemOpen
  \bibfield  {author} {\bibinfo {author} {\bibfnamefont {E.~G.}\ \bibnamefont
  {Zweibel}}\ and\ \bibinfo {author} {\bibfnamefont {H.-S.}\ \bibnamefont
  {Li}},\ }\href@noop {} {\bibfield  {journal} {\bibinfo  {journal} {Astrophys.
  J.}\ }\textbf {\bibinfo {volume} {312}},\ \bibinfo {pages} {423} (\bibinfo
  {year} {1987})}\BibitemShut {NoStop}%
\bibitem [{\citenamefont {Longcope}\ and\ \citenamefont
  {Cowley}(1996)}]{LongcopeC1996}%
  \BibitemOpen
  \bibfield  {author} {\bibinfo {author} {\bibfnamefont {D.~W.}\ \bibnamefont
  {Longcope}}\ and\ \bibinfo {author} {\bibfnamefont {S.~C.}\ \bibnamefont
  {Cowley}},\ }\href@noop {} {\bibfield  {journal} {\bibinfo  {journal} {Phys.
  Plasmas}\ }\textbf {\bibinfo {volume} {3}},\ \bibinfo {pages} {2885}
  (\bibinfo {year} {1996})}\BibitemShut {NoStop}%
\bibitem [{\citenamefont {Longbottom}\ \emph {et~al.}(1998)\citenamefont
  {Longbottom}, \citenamefont {Rickard}, \citenamefont {Craig},\ and\
  \citenamefont {Sneyd}}]{LongbottomRCS1998}%
  \BibitemOpen
  \bibfield  {author} {\bibinfo {author} {\bibfnamefont {A.~W.}\ \bibnamefont
  {Longbottom}}, \bibinfo {author} {\bibfnamefont {G.~J.}\ \bibnamefont
  {Rickard}}, \bibinfo {author} {\bibfnamefont {I.~J.~D.}\ \bibnamefont
  {Craig}},\ and\ \bibinfo {author} {\bibfnamefont {A.~D.}\ \bibnamefont
  {Sneyd}},\ }\href@noop {} {\bibfield  {journal} {\bibinfo  {journal}
  {Astrophys. J.}\ }\textbf {\bibinfo {volume} {500}},\ \bibinfo {pages} {471}
  (\bibinfo {year} {1998})}\BibitemShut {NoStop}%
\bibitem [{\citenamefont {Ng}\ and\ \citenamefont
  {Bhattacharjee}(1998)}]{NgB1998}%
  \BibitemOpen
  \bibfield  {author} {\bibinfo {author} {\bibfnamefont {C.~S.}\ \bibnamefont
  {Ng}}\ and\ \bibinfo {author} {\bibfnamefont {A.}~\bibnamefont
  {Bhattacharjee}},\ }\href@noop {} {\bibfield  {journal} {\bibinfo  {journal}
  {Phys. Plasmas}\ }\textbf {\bibinfo {volume} {5}},\ \bibinfo {pages} {4028}
  (\bibinfo {year} {1998})}\BibitemShut {NoStop}%
\bibitem [{\citenamefont {Craig}\ and\ \citenamefont
  {Sneyd}(2005)}]{CraigS2005}%
  \BibitemOpen
  \bibfield  {author} {\bibinfo {author} {\bibfnamefont {I.~J.~D.}\
  \bibnamefont {Craig}}\ and\ \bibinfo {author} {\bibfnamefont {A.~D.}\
  \bibnamefont {Sneyd}},\ }\href@noop {} {\bibfield  {journal} {\bibinfo
  {journal} {Solar Physics}\ }\textbf {\bibinfo {volume} {232}},\ \bibinfo
  {pages} {41} (\bibinfo {year} {2005})}\BibitemShut {NoStop}%
\bibitem [{\citenamefont {Low}(2006)}]{Low2006a}%
  \BibitemOpen
  \bibfield  {author} {\bibinfo {author} {\bibfnamefont {B.~C.}\ \bibnamefont
  {Low}},\ }\href@noop {} {\bibfield  {journal} {\bibinfo  {journal}
  {Astrophys. J.}\ }\textbf {\bibinfo {volume} {649}},\ \bibinfo {pages} {1064}
  (\bibinfo {year} {2006})}\BibitemShut {NoStop}%
\bibitem [{\citenamefont {Low}(2007)}]{Low2007}%
  \BibitemOpen
  \bibfield  {author} {\bibinfo {author} {\bibfnamefont {B.~C.}\ \bibnamefont
  {Low}},\ }\href@noop {} {\bibfield  {journal} {\bibinfo  {journal} {Phys.
  Plasmas}\ }\textbf {\bibinfo {volume} {14}},\ \bibinfo {pages} {122904}
  (\bibinfo {year} {2007})}\BibitemShut {NoStop}%
\bibitem [{\citenamefont {Janse}\ and\ \citenamefont {Low}(2009)}]{JanseL2009}%
  \BibitemOpen
  \bibfield  {author} {\bibinfo {author} {\bibfnamefont {A.~M.}\ \bibnamefont
  {Janse}}\ and\ \bibinfo {author} {\bibfnamefont {B.~C.}\ \bibnamefont
  {Low}},\ }\href@noop {} {\bibfield  {journal} {\bibinfo  {journal}
  {Astrophys. J.}\ }\textbf {\bibinfo {volume} {690}},\ \bibinfo {pages} {1089}
  (\bibinfo {year} {2009})}\BibitemShut {NoStop}%
\bibitem [{\citenamefont {Huang}\ \emph {et~al.}(2009)\citenamefont {Huang},
  \citenamefont {Bhattacharjee},\ and\ \citenamefont {Zweibel}}]{HuangBZ2009}%
  \BibitemOpen
  \bibfield  {author} {\bibinfo {author} {\bibfnamefont {Y.-M.}\ \bibnamefont
  {Huang}}, \bibinfo {author} {\bibfnamefont {A.}~\bibnamefont
  {Bhattacharjee}},\ and\ \bibinfo {author} {\bibfnamefont {E.~G.}\
  \bibnamefont {Zweibel}},\ }\href
  {https://doi.org/10.1088/0004-637X/699/2/L144} {\bibfield  {journal}
  {\bibinfo  {journal} {Astrophys. J. Lett.}\ }\textbf {\bibinfo {volume}
  {699}},\ \bibinfo {pages} {L144} (\bibinfo {year} {2009})}\BibitemShut
  {NoStop}%
\bibitem [{\citenamefont {Huang}\ \emph {et~al.}(2010)\citenamefont {Huang},
  \citenamefont {Bhattacharjee},\ and\ \citenamefont {Zweibel}}]{HuangBZ2010}%
  \BibitemOpen
  \bibfield  {author} {\bibinfo {author} {\bibfnamefont {Y.-M.}\ \bibnamefont
  {Huang}}, \bibinfo {author} {\bibfnamefont {A.}~\bibnamefont
  {Bhattacharjee}},\ and\ \bibinfo {author} {\bibfnamefont {E.~G.}\
  \bibnamefont {Zweibel}},\ }\href {https://doi.org/10.1063/1.3398486}
  {\bibfield  {journal} {\bibinfo  {journal} {Phys. Plasmas}\ }\textbf
  {\bibinfo {volume} {17}},\ \bibinfo {pages} {055707} (\bibinfo {year}
  {2010})}\BibitemShut {NoStop}%
\bibitem [{\citenamefont {Aly}\ and\ \citenamefont {Amari}(2010)}]{AlyA2010}%
  \BibitemOpen
  \bibfield  {author} {\bibinfo {author} {\bibfnamefont {J.~J.}\ \bibnamefont
  {Aly}}\ and\ \bibinfo {author} {\bibfnamefont {T.}~\bibnamefont {Amari}},\
  }\href@noop {} {\bibfield  {journal} {\bibinfo  {journal} {Astrophys. J.
  Lett.}\ }\textbf {\bibinfo {volume} {709}},\ \bibinfo {pages} {L99} (\bibinfo
  {year} {2010})}\BibitemShut {NoStop}%
\bibitem [{\citenamefont {Low}(2010{\natexlab{a}})}]{Low2010}%
  \BibitemOpen
  \bibfield  {author} {\bibinfo {author} {\bibfnamefont {B.~C.}\ \bibnamefont
  {Low}},\ }\href@noop {} {\bibfield  {journal} {\bibinfo  {journal}
  {Astrophys. J.}\ }\textbf {\bibinfo {volume} {718}},\ \bibinfo {pages} {717}
  (\bibinfo {year} {2010}{\natexlab{a}})}\BibitemShut {NoStop}%
\bibitem [{\citenamefont {Low}(2010{\natexlab{b}})}]{Low2010a}%
  \BibitemOpen
  \bibfield  {author} {\bibinfo {author} {\bibfnamefont {B.~C.}\ \bibnamefont
  {Low}},\ }\href@noop {} {\bibfield  {journal} {\bibinfo  {journal} {Solar
  Physics}\ }\textbf {\bibinfo {volume} {266}},\ \bibinfo {pages} {277}
  (\bibinfo {year} {2010}{\natexlab{b}})}\BibitemShut {NoStop}%
\bibitem [{\citenamefont {Janse}\ \emph {et~al.}(2010)\citenamefont {Janse},
  \citenamefont {Low},\ and\ \citenamefont {Parker}}]{JanseLP2010}%
  \BibitemOpen
  \bibfield  {author} {\bibinfo {author} {\bibfnamefont {A.~M.}\ \bibnamefont
  {Janse}}, \bibinfo {author} {\bibfnamefont {B.~C.}\ \bibnamefont {Low}},\
  and\ \bibinfo {author} {\bibfnamefont {E.~N.}\ \bibnamefont {Parker}},\
  }\href@noop {} {\bibfield  {journal} {\bibinfo  {journal} {Phys. Plasmas}\
  }\textbf {\bibinfo {volume} {17}},\ \bibinfo {pages} {092901} (\bibinfo
  {year} {2010})}\BibitemShut {NoStop}%
\bibitem [{\citenamefont {Low}(2011)}]{Low2011}%
  \BibitemOpen
  \bibfield  {author} {\bibinfo {author} {\bibfnamefont {B.~C.}\ \bibnamefont
  {Low}},\ }\href@noop {} {\bibfield  {journal} {\bibinfo  {journal} {Phys.
  Plasmas}\ }\textbf {\bibinfo {volume} {18}},\ \bibinfo {pages} {052901}
  (\bibinfo {year} {2011})}\BibitemShut {NoStop}%
\bibitem [{\citenamefont {Pontin}\ and\ \citenamefont
  {Huang}(2012)}]{PontinH2012}%
  \BibitemOpen
  \bibfield  {author} {\bibinfo {author} {\bibfnamefont {D.~I.}\ \bibnamefont
  {Pontin}}\ and\ \bibinfo {author} {\bibfnamefont {Y.-M.}\ \bibnamefont
  {Huang}},\ }\href {https://doi.org/10.1088/0004-637X/756/1/7} {\bibfield
  {journal} {\bibinfo  {journal} {Astrophys. J.}\ }\textbf {\bibinfo {volume}
  {756}},\ \bibinfo {pages} {7} (\bibinfo {year} {2012})}\BibitemShut {NoStop}%
\bibitem [{\citenamefont {Craig}\ and\ \citenamefont
  {Pontin}(2014)}]{CraigP2014}%
  \BibitemOpen
  \bibfield  {author} {\bibinfo {author} {\bibfnamefont {I.~J.~D.}\
  \bibnamefont {Craig}}\ and\ \bibinfo {author} {\bibfnamefont {D.~I.}\
  \bibnamefont {Pontin}},\ }\href {https://doi.org/10.1088/0004-637X/788/2/177}
  {\bibfield  {journal} {\bibinfo  {journal} {Astrophys. J.}\ }\textbf
  {\bibinfo {volume} {788}},\ \bibinfo {pages} {177} (\bibinfo {year}
  {2014})}\BibitemShut {NoStop}%
\bibitem [{\citenamefont {Candelaresi}\ \emph {et~al.}(2015)\citenamefont
  {Candelaresi}, \citenamefont {Pontin},\ and\ \citenamefont
  {Hornig}}]{CandelaresiPH2015}%
  \BibitemOpen
  \bibfield  {author} {\bibinfo {author} {\bibfnamefont {S.}~\bibnamefont
  {Candelaresi}}, \bibinfo {author} {\bibfnamefont {D.~I.}\ \bibnamefont
  {Pontin}},\ and\ \bibinfo {author} {\bibfnamefont {G.}~\bibnamefont
  {Hornig}},\ }\href {https://doi.org/10.1088/0004-637X/808/2/134} {\bibfield
  {journal} {\bibinfo  {journal} {Astrophys. J.}\ }\textbf {\bibinfo {volume}
  {808}},\ \bibinfo {pages} {134} (\bibinfo {year} {2015})}\BibitemShut
  {NoStop}%
\bibitem [{\citenamefont {Zhou}\ \emph {et~al.}(2018)\citenamefont {Zhou},
  \citenamefont {Huang}, \citenamefont {Qin},\ and\ \citenamefont
  {Bhattacharjee}}]{ZhouHQB2018}%
  \BibitemOpen
  \bibfield  {author} {\bibinfo {author} {\bibfnamefont {Y.}~\bibnamefont
  {Zhou}}, \bibinfo {author} {\bibfnamefont {Y.-M.}\ \bibnamefont {Huang}},
  \bibinfo {author} {\bibfnamefont {H.}~\bibnamefont {Qin}},\ and\ \bibinfo
  {author} {\bibfnamefont {A.}~\bibnamefont {Bhattacharjee}},\ }\href
  {https://doi.org/10.3847/1538-4357/aa9b84} {\bibfield  {journal} {\bibinfo
  {journal} {Astrophys. J.}\ }\textbf {\bibinfo {volume} {852}},\ \bibinfo
  {pages} {3} (\bibinfo {year} {2018})}\BibitemShut {NoStop}%
\bibitem [{\citenamefont {Pontin}\ and\ \citenamefont
  {Hornig}(2020)}]{PontinH2020}%
  \BibitemOpen
  \bibfield  {author} {\bibinfo {author} {\bibfnamefont {D.~I.}\ \bibnamefont
  {Pontin}}\ and\ \bibinfo {author} {\bibfnamefont {G.}~\bibnamefont
  {Hornig}},\ }\bibfield  {journal} {\bibinfo  {journal} {Living Reviews in
  Solar Physics}\ }\textbf {\bibinfo {volume} {17}},\ \href
  {https://doi.org/10.1007/s41116-020-00026-5} {10.1007/s41116-020-00026-5}
  (\bibinfo {year} {2020})\BibitemShut {NoStop}%
\bibitem [{Note1()}]{Note1}%
  \BibitemOpen
  \bibinfo {note} {A.~H.~Boozer, private communication (2022).}\BibitemShut
  {Stop}%
\bibitem [{\citenamefont {Kadomtsev}\ and\ \citenamefont
  {Pogutse}(1974)}]{KadomtsevP1974}%
  \BibitemOpen
  \bibfield  {author} {\bibinfo {author} {\bibfnamefont {B.~B.}\ \bibnamefont
  {Kadomtsev}}\ and\ \bibinfo {author} {\bibfnamefont {O.~P.}\ \bibnamefont
  {Pogutse}},\ }\href@noop {} {\bibfield  {journal} {\bibinfo  {journal} {Sov.
  Phys. JETP}\ }\textbf {\bibinfo {volume} {38}},\ \bibinfo {pages} {283}
  (\bibinfo {year} {1974})}\BibitemShut {NoStop}%
\bibitem [{\citenamefont {Strauss}(1976)}]{Strauss1976}%
  \BibitemOpen
  \bibfield  {author} {\bibinfo {author} {\bibfnamefont {H.~R.}\ \bibnamefont
  {Strauss}},\ }\href@noop {} {\bibfield  {journal} {\bibinfo  {journal} {Phys.
  Fluids}\ }\textbf {\bibinfo {volume} {19}},\ \bibinfo {pages} {134} (\bibinfo
  {year} {1976})}\BibitemShut {NoStop}%
\bibitem [{\citenamefont {Strauss}\ and\ \citenamefont
  {Otani}(1988)}]{StraussO1988}%
  \BibitemOpen
  \bibfield  {author} {\bibinfo {author} {\bibfnamefont {H.~R.}\ \bibnamefont
  {Strauss}}\ and\ \bibinfo {author} {\bibfnamefont {N.~F.}\ \bibnamefont
  {Otani}},\ }\href@noop {} {\bibfield  {journal} {\bibinfo  {journal}
  {Astrophys. J.}\ }\textbf {\bibinfo {volume} {326}},\ \bibinfo {pages} {418}
  (\bibinfo {year} {1988})}\BibitemShut {NoStop}%
\bibitem [{\citenamefont {Longcope}\ and\ \citenamefont
  {Strauss}(1994)}]{LongcopeS1994a}%
  \BibitemOpen
  \bibfield  {author} {\bibinfo {author} {\bibfnamefont {D.~W.}\ \bibnamefont
  {Longcope}}\ and\ \bibinfo {author} {\bibfnamefont {H.~R.}\ \bibnamefont
  {Strauss}},\ }\href@noop {} {\bibfield  {journal} {\bibinfo  {journal}
  {Astrophys. J.}\ }\textbf {\bibinfo {volume} {437}},\ \bibinfo {pages} {851}
  (\bibinfo {year} {1994})}\BibitemShut {NoStop}%
\bibitem [{\citenamefont {Longcope}\ and\ \citenamefont
  {Sudan}(1994)}]{LongcopeS1994b}%
  \BibitemOpen
  \bibfield  {author} {\bibinfo {author} {\bibfnamefont {D.~W.}\ \bibnamefont
  {Longcope}}\ and\ \bibinfo {author} {\bibfnamefont {R.~N.}\ \bibnamefont
  {Sudan}},\ }\href@noop {} {\bibfield  {journal} {\bibinfo  {journal}
  {Astrophys. J.}\ }\textbf {\bibinfo {volume} {437}},\ \bibinfo {pages} {491}
  (\bibinfo {year} {1994})}\BibitemShut {NoStop}%
\bibitem [{\citenamefont {Dmitruk}\ and\ \citenamefont
  {G\'omez}(1999)}]{DmitrukG1999}%
  \BibitemOpen
  \bibfield  {author} {\bibinfo {author} {\bibfnamefont {P.}~\bibnamefont
  {Dmitruk}}\ and\ \bibinfo {author} {\bibfnamefont {D.~O.}\ \bibnamefont
  {G\'omez}},\ }\href@noop {} {\bibfield  {journal} {\bibinfo  {journal}
  {Astrophys. J.}\ }\textbf {\bibinfo {volume} {527}},\ \bibinfo {pages} {L63}
  (\bibinfo {year} {1999})}\BibitemShut {NoStop}%
\bibitem [{\citenamefont {Dmitruk}\ \emph {et~al.}(2003)\citenamefont
  {Dmitruk}, \citenamefont {G\'omez},\ and\ \citenamefont
  {Matthaeus}}]{DmitrukGM2003}%
  \BibitemOpen
  \bibfield  {author} {\bibinfo {author} {\bibfnamefont {P.}~\bibnamefont
  {Dmitruk}}, \bibinfo {author} {\bibfnamefont {D.~O.}\ \bibnamefont
  {G\'omez}},\ and\ \bibinfo {author} {\bibfnamefont {W.~H.}\ \bibnamefont
  {Matthaeus}},\ }\href@noop {} {\bibfield  {journal} {\bibinfo  {journal}
  {Phys. Plasmas}\ }\textbf {\bibinfo {volume} {10}},\ \bibinfo {pages} {3584}
  (\bibinfo {year} {2003})}\BibitemShut {NoStop}%
\bibitem [{\citenamefont {Rappazzo}\ \emph {et~al.}(2007)\citenamefont
  {Rappazzo}, \citenamefont {Velli}, \citenamefont {Einaudi},\ and\
  \citenamefont {Dahlburg}}]{RappazzoVED2007}%
  \BibitemOpen
  \bibfield  {author} {\bibinfo {author} {\bibfnamefont {A.~F.}\ \bibnamefont
  {Rappazzo}}, \bibinfo {author} {\bibfnamefont {M.}~\bibnamefont {Velli}},
  \bibinfo {author} {\bibfnamefont {G.}~\bibnamefont {Einaudi}},\ and\ \bibinfo
  {author} {\bibfnamefont {R.~B.}\ \bibnamefont {Dahlburg}},\ }\href@noop {}
  {\bibfield  {journal} {\bibinfo  {journal} {Astrophys. J.}\ }\textbf
  {\bibinfo {volume} {657}},\ \bibinfo {pages} {L47} (\bibinfo {year}
  {2007})}\BibitemShut {NoStop}%
\bibitem [{\citenamefont {Rappazzo}\ \emph {et~al.}(2008)\citenamefont
  {Rappazzo}, \citenamefont {Velli}, \citenamefont {Einaudi},\ and\
  \citenamefont {Dahlburg}}]{RappazzoVED2008}%
  \BibitemOpen
  \bibfield  {author} {\bibinfo {author} {\bibfnamefont {A.~F.}\ \bibnamefont
  {Rappazzo}}, \bibinfo {author} {\bibfnamefont {M.}~\bibnamefont {Velli}},
  \bibinfo {author} {\bibfnamefont {G.}~\bibnamefont {Einaudi}},\ and\ \bibinfo
  {author} {\bibfnamefont {R.~B.}\ \bibnamefont {Dahlburg}},\ }\href@noop {}
  {\bibfield  {journal} {\bibinfo  {journal} {Astrophys. J.}\ }\textbf
  {\bibinfo {volume} {677}},\ \bibinfo {pages} {1348} (\bibinfo {year}
  {2008})}\BibitemShut {NoStop}%
\bibitem [{\citenamefont {Ng}\ and\ \citenamefont
  {Bhattacharjee}(2008)}]{NgB2008}%
  \BibitemOpen
  \bibfield  {author} {\bibinfo {author} {\bibfnamefont {C.~S.}\ \bibnamefont
  {Ng}}\ and\ \bibinfo {author} {\bibfnamefont {A.}~\bibnamefont
  {Bhattacharjee}},\ }\href@noop {} {\bibfield  {journal} {\bibinfo  {journal}
  {Astrophys. J.}\ }\textbf {\bibinfo {volume} {675}},\ \bibinfo {pages} {899}
  (\bibinfo {year} {2008})}\BibitemShut {NoStop}%
\bibitem [{\citenamefont {Ng}\ \emph {et~al.}(2012)\citenamefont {Ng},
  \citenamefont {Lin},\ and\ \citenamefont {Bhattacharjee}}]{NgLB2012}%
  \BibitemOpen
  \bibfield  {author} {\bibinfo {author} {\bibfnamefont {C.~S.}\ \bibnamefont
  {Ng}}, \bibinfo {author} {\bibfnamefont {L.}~\bibnamefont {Lin}},\ and\
  \bibinfo {author} {\bibfnamefont {A.}~\bibnamefont {Bhattacharjee}},\
  }\href@noop {} {\bibfield  {journal} {\bibinfo  {journal} {Astrophys. J.}\
  }\textbf {\bibinfo {volume} {747}},\ \bibinfo {pages} {109} (\bibinfo {year}
  {2012})}\BibitemShut {NoStop}%
\bibitem [{\citenamefont {Rappazzo}\ and\ \citenamefont
  {Parker}(2013)}]{RappazzoP2013}%
  \BibitemOpen
  \bibfield  {author} {\bibinfo {author} {\bibfnamefont {A.~F.}\ \bibnamefont
  {Rappazzo}}\ and\ \bibinfo {author} {\bibfnamefont {E.~N.}\ \bibnamefont
  {Parker}},\ }\href {https://doi.org/10.1088/2041-8205/773/1/L2} {\bibfield
  {journal} {\bibinfo  {journal} {Astrophys. J. Lett.}\ }\textbf {\bibinfo
  {volume} {773}},\ \bibinfo {pages} {L2} (\bibinfo {year} {2013})}\BibitemShut
  {NoStop}%
\bibitem [{\citenamefont {Huang}\ \emph {et~al.}(2014)\citenamefont {Huang},
  \citenamefont {Bhattacharjee},\ and\ \citenamefont {Boozer}}]{HuangBB2014}%
  \BibitemOpen
  \bibfield  {author} {\bibinfo {author} {\bibfnamefont {Y.-M.}\ \bibnamefont
  {Huang}}, \bibinfo {author} {\bibfnamefont {A.}~\bibnamefont
  {Bhattacharjee}},\ and\ \bibinfo {author} {\bibfnamefont {A.~H.}\
  \bibnamefont {Boozer}},\ }\href {https://doi.org/10.1088/0004-637X/793/2/106}
  {\bibfield  {journal} {\bibinfo  {journal} {Astrophys. J.}\ }\textbf
  {\bibinfo {volume} {793}},\ \bibinfo {pages} {106} (\bibinfo {year}
  {2014})}\BibitemShut {NoStop}%
\bibitem [{\citenamefont {Schnack}\ \emph {et~al.}(1986)\citenamefont
  {Schnack}, \citenamefont {Barnes}, \citenamefont {Miki\'c}, \citenamefont
  {Harned}, \citenamefont {Caramana},\ and\ \citenamefont
  {Nebel}}]{SchnackBMHCN1986}%
  \BibitemOpen
  \bibfield  {author} {\bibinfo {author} {\bibfnamefont {D.~D.}\ \bibnamefont
  {Schnack}}, \bibinfo {author} {\bibfnamefont {D.~C.}\ \bibnamefont {Barnes}},
  \bibinfo {author} {\bibfnamefont {Z.}~\bibnamefont {Miki\'c}}, \bibinfo
  {author} {\bibfnamefont {D.~S.}\ \bibnamefont {Harned}}, \bibinfo {author}
  {\bibfnamefont {E.~J.}\ \bibnamefont {Caramana}},\ and\ \bibinfo {author}
  {\bibfnamefont {R.~A.}\ \bibnamefont {Nebel}},\ }\href@noop {} {\bibfield
  {journal} {\bibinfo  {journal} {Computer Physics Communications}\ }\textbf
  {\bibinfo {volume} {43}},\ \bibinfo {pages} {17} (\bibinfo {year}
  {1986})}\BibitemShut {NoStop}%
\bibitem [{\citenamefont {Courant}\ \emph {et~al.}(1928)\citenamefont
  {Courant}, \citenamefont {Friedrichs},\ and\ \citenamefont
  {Lewy}}]{CourantFL1928}%
  \BibitemOpen
  \bibfield  {author} {\bibinfo {author} {\bibfnamefont {R.}~\bibnamefont
  {Courant}}, \bibinfo {author} {\bibfnamefont {K.}~\bibnamefont
  {Friedrichs}},\ and\ \bibinfo {author} {\bibfnamefont {H.}~\bibnamefont
  {Lewy}},\ }\href {https://doi.org/10.1007/bf01448839} {\bibfield  {journal}
  {\bibinfo  {journal} {Mathematische Annalen}\ }\textbf {\bibinfo {volume}
  {100}},\ \bibinfo {pages} {32} (\bibinfo {year} {1928})}\BibitemShut
  {NoStop}%
\bibitem [{\citenamefont {Arfken}\ \emph {et~al.}(2013)\citenamefont {Arfken},
  \citenamefont {Weber},\ and\ \citenamefont {Harris}}]{ArfkenWH2013}%
  \BibitemOpen
  \bibfield  {author} {\bibinfo {author} {\bibfnamefont {G.~B.}\ \bibnamefont
  {Arfken}}, \bibinfo {author} {\bibfnamefont {H.~J.}\ \bibnamefont {Weber}},\
  and\ \bibinfo {author} {\bibfnamefont {F.~E.}\ \bibnamefont {Harris}},\
  }\href@noop {} {\emph {\bibinfo {title} {Mathematical Methods for Physicists:
  A Comprehensive Guide}}},\ \bibinfo {edition} {seventh}\ ed.\ (\bibinfo
  {publisher} {Academic Press},\ \bibinfo {year} {2013})\BibitemShut {NoStop}%
\bibitem [{Note2()}]{Note2}%
  \BibitemOpen
  \bibinfo {note} {The eigenfunction expansion should include the contribution
  of continuous spectra if they exist. For continuous spectra, the summation in
  Eq.~(\ref {eq:eigen_expansion}) should be replaced by an integral. However,
  continuous spectra are often associated with toroidal magnetic fields with
  nested flux surfaces. They are likely to be absent in the coronal loop model
  of this study due to the line-tied boundary condition.}\BibitemShut {Stop}%
\bibitem [{Note3()}]{Note3}%
  \BibitemOpen
  \bibinfo {note} {The derivation of the BE equation given by Boozer and Elder
  is slightly different from ours. They start by writing Eq.~(\ref
  {eq:RMHD-faraday})(with $\eta =0$) in the Lagrangian coordinates as $\left
  (\partial _{t}A\right )_{L}=\left (\partial _{z}\phi \right )_{L}$. Here, the
  subscript $L$ implies the use of Lagrangian coordinates. Next, they apply
  $-\nabla _{\perp }^{2}$ (also expressed in Lagrangian coordinates) on both
  sides and obtain $\left (\partial _{t}J\right )_{L}=\left (\partial
  _{z}\Omega \right )_{L}$, which is equivalent to Eq.~(\ref {eq:BE1}).
  However, this derivation neglects the fact that the Laplacian $\nabla _{\perp
  }^{2}$ has explicit time-dependence when it is expressed in Lagrangian
  coordinates because the Lagrangian coordinates themselves are time-dependent.
  As such, the Laplacian $\nabla _{\perp }^{2}$ and the time derivative
  $\partial _{t}$ do not commute. Consequently, $\left (\partial _{t}J\right
  )_{L}=\left (-\partial _{t}\nabla _{\perp }^{2}A\right )\protect \neq \left
  (-\nabla _{\perp }^{2}\partial _{t}A\right )_{L}$, implying some terms are
  missing in their derivation. This same issue of missing terms also appears in
  some other publications, e.g., Eqs.~(D4) and (D5) of Ref. {[}\protect
  \rev@citealpnum {Boozer2018}{]} and Eq.~(52) of Ref. {[}\protect
  \rev@citealpnum {Boozer2022}{]}.}\BibitemShut {Stop}%
\bibitem [{\citenamefont {Burns}\ \emph {et~al.}(2020)\citenamefont {Burns},
  \citenamefont {Vasil}, \citenamefont {Oishi}, \citenamefont {Lecoanet},\ and\
  \citenamefont {Brown}}]{BurnsVOLB2020}%
  \BibitemOpen
  \bibfield  {author} {\bibinfo {author} {\bibfnamefont {K.~J.}\ \bibnamefont
  {Burns}}, \bibinfo {author} {\bibfnamefont {G.~M.}\ \bibnamefont {Vasil}},
  \bibinfo {author} {\bibfnamefont {J.~S.}\ \bibnamefont {Oishi}}, \bibinfo
  {author} {\bibfnamefont {D.}~\bibnamefont {Lecoanet}},\ and\ \bibinfo
  {author} {\bibfnamefont {B.~P.}\ \bibnamefont {Brown}},\ }\href
  {https://doi.org/10.1103/physrevresearch.2.023068} {\bibfield  {journal}
  {\bibinfo  {journal} {Physical Review Research}\ }\textbf {\bibinfo {volume}
  {2}},\ \bibinfo {pages} {023068} (\bibinfo {year} {2020})}\BibitemShut
  {NoStop}%
\bibitem [{\citenamefont {Boyd}(2001)}]{Boyd2001}%
  \BibitemOpen
  \bibfield  {author} {\bibinfo {author} {\bibfnamefont {J.~P.}\ \bibnamefont
  {Boyd}},\ }\href@noop {} {\emph {\bibinfo {title} {Chebyshev and Fourier
  Spectral Methods}}},\ \bibinfo {edition} {2nd}\ ed.\ (\bibinfo  {publisher}
  {Dover Publications, Inc.},\ \bibinfo {year} {2001})\BibitemShut {NoStop}%
\bibitem [{\citenamefont {Sweet}(1958)}]{Sweet1958a}%
  \BibitemOpen
  \bibfield  {author} {\bibinfo {author} {\bibfnamefont {P.~A.}\ \bibnamefont
  {Sweet}},\ }in\ \href@noop {} {\emph {\bibinfo {booktitle} {Electromagnetic
  phenomena in cosmical physics}}},\ Vol.~\bibinfo {volume} {6},\ \bibinfo
  {editor} {edited by\ \bibinfo {editor} {\bibfnamefont {B.}~\bibnamefont
  {Lehnert}}}\ (\bibinfo  {publisher} {Cambridge University Press},\ \bibinfo
  {year} {1958})\ pp.\ \bibinfo {pages} {123--134}\BibitemShut {NoStop}%
\bibitem [{\citenamefont {Parker}(1957)}]{Parker1957}%
  \BibitemOpen
  \bibfield  {author} {\bibinfo {author} {\bibfnamefont {E.~N.}\ \bibnamefont
  {Parker}},\ }\href {https://doi.org/10.1029/JZ062i004p00509} {\bibfield
  {journal} {\bibinfo  {journal} {J. Geophys. Res.}\ }\textbf {\bibinfo
  {volume} {62}},\ \bibinfo {pages} {509} (\bibinfo {year} {1957})}\BibitemShut
  {NoStop}%
\bibitem [{\citenamefont {Huang}\ and\ \citenamefont
  {Bhattacharjee}(2010)}]{HuangB2010}%
  \BibitemOpen
  \bibfield  {author} {\bibinfo {author} {\bibfnamefont {Y.-M.}\ \bibnamefont
  {Huang}}\ and\ \bibinfo {author} {\bibfnamefont {A.}~\bibnamefont
  {Bhattacharjee}},\ }\href {https://doi.org/10.1063/1.3420208} {\bibfield
  {journal} {\bibinfo  {journal} {Phys. Plasmas}\ }\textbf {\bibinfo {volume}
  {17}},\ \bibinfo {pages} {062104} (\bibinfo {year} {2010})}\BibitemShut
  {NoStop}%
\bibitem [{\citenamefont {Bhattacharjee}\ \emph {et~al.}(2009)\citenamefont
  {Bhattacharjee}, \citenamefont {Huang}, \citenamefont {Yang},\ and\
  \citenamefont {Rogers}}]{BhattacharjeeHYR2009}%
  \BibitemOpen
  \bibfield  {author} {\bibinfo {author} {\bibfnamefont {A.}~\bibnamefont
  {Bhattacharjee}}, \bibinfo {author} {\bibfnamefont {Y.-M.}\ \bibnamefont
  {Huang}}, \bibinfo {author} {\bibfnamefont {H.}~\bibnamefont {Yang}},\ and\
  \bibinfo {author} {\bibfnamefont {B.}~\bibnamefont {Rogers}},\ }\href
  {https://doi.org/10.1063/1.3264103} {\bibfield  {journal} {\bibinfo
  {journal} {Phys. Plasmas}\ }\textbf {\bibinfo {volume} {16}},\ \bibinfo
  {pages} {112102} (\bibinfo {year} {2009})}\BibitemShut {NoStop}%
\bibitem [{\citenamefont {Huang}\ \emph {et~al.}(2017)\citenamefont {Huang},
  \citenamefont {Comisso},\ and\ \citenamefont {Bhattacharjee}}]{HuangCB2017}%
  \BibitemOpen
  \bibfield  {author} {\bibinfo {author} {\bibfnamefont {Y.-M.}\ \bibnamefont
  {Huang}}, \bibinfo {author} {\bibfnamefont {L.}~\bibnamefont {Comisso}},\
  and\ \bibinfo {author} {\bibfnamefont {A.}~\bibnamefont {Bhattacharjee}},\
  }\href {https://doi.org/10.3847/1538-4357/aa906d} {\bibfield  {journal}
  {\bibinfo  {journal} {Astrophys. J.}\ }\textbf {\bibinfo {volume} {849}},\
  \bibinfo {pages} {75} (\bibinfo {year} {2017})}\BibitemShut {NoStop}%
\bibitem [{\citenamefont {Huang}\ \emph {et~al.}(2019)\citenamefont {Huang},
  \citenamefont {Comisso},\ and\ \citenamefont {Bhattacharjee}}]{HuangCB2019}%
  \BibitemOpen
  \bibfield  {author} {\bibinfo {author} {\bibfnamefont {Y.-M.}\ \bibnamefont
  {Huang}}, \bibinfo {author} {\bibfnamefont {L.}~\bibnamefont {Comisso}},\
  and\ \bibinfo {author} {\bibfnamefont {A.}~\bibnamefont {Bhattacharjee}},\
  }\href {https://doi.org/10.1063/1.5110332} {\bibfield  {journal} {\bibinfo
  {journal} {Physics of Plasmas}\ }\textbf {\bibinfo {volume} {26}},\ \bibinfo
  {pages} {092112} (\bibinfo {year} {2019})}\BibitemShut {NoStop}%
\bibitem [{\citenamefont {Delzanno}\ and\ \citenamefont
  {Finn}(2008)}]{DelzannoF2008}%
  \BibitemOpen
  \bibfield  {author} {\bibinfo {author} {\bibfnamefont {G.~L.}\ \bibnamefont
  {Delzanno}}\ and\ \bibinfo {author} {\bibfnamefont {J.~M.}\ \bibnamefont
  {Finn}},\ }\href@noop {} {\bibfield  {journal} {\bibinfo  {journal} {Phys.
  Plasmas}\ }\textbf {\bibinfo {volume} {15}},\ \bibinfo {pages} {032904}
  (\bibinfo {year} {2008})}\BibitemShut {NoStop}%
\bibitem [{\citenamefont {Huang}\ and\ \citenamefont
  {Zweibel}(2009)}]{HuangZ2009}%
  \BibitemOpen
  \bibfield  {author} {\bibinfo {author} {\bibfnamefont {Y.-M.}\ \bibnamefont
  {Huang}}\ and\ \bibinfo {author} {\bibfnamefont {E.~G.}\ \bibnamefont
  {Zweibel}},\ }\href {https://doi.org/10.1063/1.3103789} {\bibfield  {journal}
  {\bibinfo  {journal} {Physics of Plasmas}\ }\textbf {\bibinfo {volume}
  {16}},\ \bibinfo {pages} {042102} (\bibinfo {year} {2009})}\BibitemShut
  {NoStop}%
\bibitem [{\citenamefont {Richardson}\ and\ \citenamefont
  {Finn}(2012)}]{RichardsonF2012}%
  \BibitemOpen
  \bibfield  {author} {\bibinfo {author} {\bibfnamefont {A.}~\bibnamefont
  {Richardson}}\ and\ \bibinfo {author} {\bibfnamefont {J.}~\bibnamefont
  {Finn}},\ }\href
  {https://doi.org/https://doi.org/10.1016/j.cnsns.2011.04.029} {\bibfield
  {journal} {\bibinfo  {journal} {Communications in Nonlinear Science and
  Numerical Simulation}\ }\textbf {\bibinfo {volume} {17}},\ \bibinfo {pages}
  {2132} (\bibinfo {year} {2012})},\ \bibinfo {note} {special Issue:
  Mathematical Structure of Fluids and Plasmas}\BibitemShut {NoStop}%
\bibitem [{\citenamefont {Finn}\ \emph {et~al.}(2014)\citenamefont {Finn},
  \citenamefont {Billey}, \citenamefont {Daughton},\ and\ \citenamefont
  {Zweibel}}]{FinnBDZ2014}%
  \BibitemOpen
  \bibfield  {author} {\bibinfo {author} {\bibfnamefont {J.~M.}\ \bibnamefont
  {Finn}}, \bibinfo {author} {\bibfnamefont {Z.}~\bibnamefont {Billey}},
  \bibinfo {author} {\bibfnamefont {W.}~\bibnamefont {Daughton}},\ and\
  \bibinfo {author} {\bibfnamefont {E.}~\bibnamefont {Zweibel}},\ }\href
  {https://doi.org/10.1088/0741-3335/56/6/064013} {\bibfield  {journal}
  {\bibinfo  {journal} {Plasma Phys. Control. Fusion}\ }\textbf {\bibinfo
  {volume} {56}},\ \bibinfo {pages} {064013} (\bibinfo {year}
  {2014})}\BibitemShut {NoStop}%
\bibitem [{\citenamefont {Trefethen}\ and\ \citenamefont
  {Bau}(1997)}]{TrefethenB1997}%
  \BibitemOpen
  \bibfield  {author} {\bibinfo {author} {\bibfnamefont {L.~N.}\ \bibnamefont
  {Trefethen}}\ and\ \bibinfo {author} {\bibfnamefont {D.}~\bibnamefont {Bau},
  \bibfnamefont {III}},\ }\href@noop {} {\emph {\bibinfo {title} {Numerical
  Linear Algebra}}}\ (\bibinfo  {publisher} {SIAM Philadelphia},\ \bibinfo
  {year} {1997})\BibitemShut {NoStop}%
\bibitem [{Note4()}]{Note4}%
  \BibitemOpen
  \bibinfo {note} {Boozer and Elder employ the Frobenius norm of $N$, defined
  as $\left \Vert N\right \Vert =\protect \sqrt
  {N_{xx}^{2}+N_{xy}^{2}+N_{yx}^{2}+N_{yy}^{2}}$, to characterize the
  neighboring field line separation. The Frobenius norm and the squashing
  factor are related by $\left \Vert N\right \Vert =\protect \sqrt {Q}$.
  Therefore, we can calculate $Q$ without invoking SVD.}\BibitemShut {Stop}%
\bibitem [{Note5()}]{Note5}%
  \BibitemOpen
  \bibinfo {note} {Note that even though the existence of intense current
  sheets is not a necessary condition for a high squashing factor $Q$, the
  former naturally leads to the later because of the strongly sheared magnetic
  field associated with thin current sheets.}\BibitemShut {Stop}%
\bibitem [{\citenamefont {Zweibel}\ and\ \citenamefont
  {Boozer}(1985)}]{ZweibelB1985}%
  \BibitemOpen
  \bibfield  {author} {\bibinfo {author} {\bibfnamefont {E.~G.}\ \bibnamefont
  {Zweibel}}\ and\ \bibinfo {author} {\bibfnamefont {A.~H.}\ \bibnamefont
  {Boozer}},\ }\href@noop {} {\bibfield  {journal} {\bibinfo  {journal}
  {Astrophys. J.}\ }\textbf {\bibinfo {volume} {295}},\ \bibinfo {pages} {642}
  (\bibinfo {year} {1985})}\BibitemShut {NoStop}%
\bibitem [{\citenamefont {Huang}\ \emph {et~al.}(2006)\citenamefont {Huang},
  \citenamefont {Zweibel},\ and\ \citenamefont {Sovinec}}]{HuangZS2006}%
  \BibitemOpen
  \bibfield  {author} {\bibinfo {author} {\bibfnamefont {Y.-M.}\ \bibnamefont
  {Huang}}, \bibinfo {author} {\bibfnamefont {E.~G.}\ \bibnamefont {Zweibel}},\
  and\ \bibinfo {author} {\bibfnamefont {C.~R.}\ \bibnamefont {Sovinec}},\
  }\href {https://doi.org/10.1063/1.2336506} {\bibfield  {journal} {\bibinfo
  {journal} {Phys. Plasmas}\ }\textbf {\bibinfo {volume} {13}},\ \bibinfo
  {pages} {092102} (\bibinfo {year} {2006})}\BibitemShut {NoStop}%
\bibitem [{\citenamefont {Bhattacharjee}\ and\ \citenamefont
  {Wang}(1991)}]{BhattacharjeeW1991}%
  \BibitemOpen
  \bibfield  {author} {\bibinfo {author} {\bibfnamefont {A.}~\bibnamefont
  {Bhattacharjee}}\ and\ \bibinfo {author} {\bibfnamefont {X.}~\bibnamefont
  {Wang}},\ }\href@noop {} {\bibfield  {journal} {\bibinfo  {journal}
  {Astrophys. J.}\ }\textbf {\bibinfo {volume} {372}},\ \bibinfo {pages} {321}
  (\bibinfo {year} {1991})}\BibitemShut {NoStop}%
\bibitem [{\citenamefont {Boozer}(2022{\natexlab{b}})}]{Boozer2022a}%
  \BibitemOpen
  \bibfield  {author} {\bibinfo {author} {\bibfnamefont {A.~H.}\ \bibnamefont
  {Boozer}},\ }\href@noop {} {\bibinfo {title} {Judgment of paradigms for
  magnetic reconnection in coronal loops}} (\bibinfo {year}
  {2022}{\natexlab{b}}),\ \Eprint {https://arxiv.org/abs/arXiv:2210.02209}
  {arXiv:2210.02209} \BibitemShut {NoStop}%
\bibitem [{\citenamefont {Aschwanden}(2005)}]{Aschwanden2005}%
  \BibitemOpen
  \bibfield  {author} {\bibinfo {author} {\bibfnamefont {M.~J.}\ \bibnamefont
  {Aschwanden}},\ }\href@noop {} {\emph {\bibinfo {title} {Physics of the Solar
  Corona}}}\ (\bibinfo  {publisher} {Springer},\ \bibinfo {year}
  {2005})\BibitemShut {NoStop}%
\end{thebibliography}%

\end{document}